\documentclass[%
reprint,
 amsmath,amssymb,
 aps,
pra,
floatfix
]{revtex4-2}
\usepackage{graphicx}
\usepackage{dcolumn}
\usepackage{makecell}
\usepackage{bm}
\usepackage{physics}
\renewcommand{\arraystretch}{2}
\usepackage{xfrac}
\usepackage[per-mode=symbol,range-phrase = \text{--}]{siunitx}

\usepackage{newtxtext} 
\usepackage{newtxmath} 

\DeclareSymbolFont{LMletters}{OMS}{lmsy}{m}{n}
\DeclareMathAlphabet{\lmcal}{OMS}{lmsy}{m}{n}

\usepackage{enumitem}

\protected\def\vtsp{%
  \ifmmode
    \mskip0.3\thinmuskip
  \else
    \ifhmode
      \kern0.050004em
    \fi
  \fi
}

\usepackage[normalem]{ulem} 
\usepackage{xcolor}

\newlength{\oldbelowdisplayskip}
\newlength{\oldbelowdisplayshortskip}
\newlength{\oldabovedisplayskip}
\newlength{\oldabovedisplayshortskip}

\newcommand{\zeroskips}{%
\setlength{\belowdisplayskip}{0pt} \setlength{\belowdisplayshortskip}{0pt}
\setlength{\abovedisplayskip}{0pt} \setlength{\abovedisplayshortskip}{0pt}
}

\newcommand{\restoreskips}{%
\setlength{\belowdisplayskip}{\oldbelowdisplayskip} \setlength{\belowdisplayshortskip}{\oldbelowdisplayshortskip}
\setlength{\abovedisplayskip}{\oldabovedisplayskip} \setlength{\abovedisplayshortskip}{\oldabovedisplayshortskip}
}

\usepackage[defaultlines=2,all]{nowidow}

\DeclareFontFamily{OML}{pseudofourier}{}
\DeclareFontShape{OML}{pseudofourier}{m}{it}{<-> futmii}{}
\DeclareSymbolFont{pseudofouriersym}{OML}{pseudofourier}{m}{it}
\DeclareMathSymbol{\altChi}{\mathalpha}{pseudofouriersym}{31} 

\newcommand{\eqnm}[1]{Eq.~\eqref{#1}}

\begin{document}

\title{A canonical approach to quantum fluctuations}

\author{Joanna Ruhl}
\email{joanna.ruhl001@umb.edu}
 \affiliation{Department of Physics, University of Massachusetts Boston, Boston, Massachusetts 02125, USA}
 
\author{Vanja Dunjko}%
 \affiliation{Department of Physics, University of Massachusetts Boston, Boston, Massachusetts 02125, USA}

\author{Maxim Olshanii}
\affiliation{Department of Physics, University of Massachusetts Boston, Boston, Massachusetts 02125, USA}%

\date{\today}

\begin{abstract}
We present a canonical formalism for computing quantum fluctuations of certain discrete degrees of freedom in systems governed by integrable partial differential equations with known Hamiltonian structure, provided these models are classical-field approximations of underlying many-body quantum systems. We then apply the formalism to both the 2-soliton and 3-soliton breather solutions of the nonlinear Schr{\"o}dinger equation, assuming the breathers are created from an initial elementary soliton by quenching the coupling constant. In particular, we compute the immediate post-quench quantum fluctuations in the positions, velocities, norms, and phases of the constituent solitons. For each case, we consider both the white-noise and correlated-noise models for the fluctuation vacuum state. Unlike previous treatments of the problem, our method allows for analytic solutions. Additionally, in the correlated-noise case, we consider the particle-number-conserving (also called $U(1)$-symmetry-conserving) Bogoliubov modes, i.e., modes with the proper correction to preserve the total particle number. We find that in most (but not all) cases, these corrections do not change the final result.
\end{abstract}

\maketitle


\section{\label{sec:Intro}Introduction}
Integrable classical field theories are much-studied for both their conceptual and experimental importance. The conceptual importance is due to their exact solvability, most famously by the inverse scattering method \cite{Gardner1967_1095,Lax_1968,Ablowitz_PhysRevLett.30.1262,Ablowitz1974_249,Ablowitz1981,Dodd1982solitons,Novikov1984,takhtajan_book2007}. On the experimental side, many of these theories describe systems found in laboratories, nature, or both. Notably, many of these systems support soliton solutions \cite{Ablowitz1981,Dodd1982solitons,takhtajan_book2007}. The three best-known examples of theories with solitons include:
1.\ the Korteweg--de Vries (KdV) equation, which describes, e.g., long waves on the surface of shallow water, long acoustic waves in a plasma composed of cold ions and hot electrons, and nonlinear Rossby waves in the atmosphere \cite{Zab_PhysRevLett.15.240,Dodd1982solitons};
2.\ the sine-Gordon equation, which can describe, e.g., fluxon propagation in long Josephson junctions \cite{Mazo2014}, self-induced transparency in nonlinear optics, dynamics of crystal dislocations, Bloch wall dynamics in ferromagnetics and ferroelectrics, and spin waves in the A-phase of liquid \textsuperscript{3}He \cite{Ablowitz1981,Bullough2005_840};
3.\ the one-dimensional (1D) nonlinear Schr\"odinger equation (NLSE), whose applications include nonlinear optics \cite{Hasegawa1973,Lai1989a,Lai1989b,Haus1996}, small-amplitude gravity waves on the surface of deep inviscid water \cite{Zakharov1968_86}, Langmuir waves in hot plasmas \cite{malomed2005_639}, Davydov solitons in $\alpha$-helix proteins \cite{Davydov1990_11}, and Bose-Einstein condensates (BECs) \cite{%
Burger1999_5198, 
Denschlag2000_97, 
Strecker2002_150_Nature_with_London, 
Khaykovich2002, 
Strecker2003, 
Cornish2006, 
Marchant2013, 
Di_Carli2019_123602, 
Luo2020_183902}. 



When applied to experiments or nature, each of these theories is, of course, an approximation, albeit an excellent one, with a well-understood domain of validity. Under certain circumstances, however, even while formally remaining within its domain of validity, it may be possible to detect the effects of the more exact theory that underlies the classical field theory. 

As a concrete example, consider the NLSE as it applies to ultracold atomic gases. A gas of bosons forms a BEC when it is cooled to the point that the de Broglie wavelength of the atoms becomes comparable to the interparticle distance. One expects---and all available experimental evidence confirms---that all experimentally accessible properties of such a system are accounted for by the nonrelativistic quantum many-body physics \cite{Leggett2001_307,Bloch2008_885}. Moreover, the parameter regime is often such that the mean-field (i.e., classical field) description is correct to an excellent approximation. The simplest way to justify this is via the Hartree approximation (within which a macroscopic number of atoms occupy the single quantum state of lowest energy) \cite{Leggett2001_307}.

If the BEC is further placed in a highly anisotropic cigar-shaped trap, then the classical 1D NLSE provides a very good mean-field description of the gas \cite{Castin2001_419}. 

At the next level of approximation, a widely used prescription for describing small quantum excitations is the linearization method proposed by Bogoliubov for superfluid quantum liquids \cite{Bogolubov1947}, which has been applied extensively to ultracold gases \cite{Griffin1995,Dalfovo1999,Leggett2001_307,Pethick2008}. Of special relevance to us is that it was also applied to fundamental solitons in optical fibers \cite{Haus1990,Lai1993} and then to center-of-mass degrees of freedom in soliton breathers \cite{Yeang1999}. 

In a recent work with our collaborators \cite{marchukov2020}, we applied the Bogoliubov method to compute the quantum fluctuations of relative macroscopic parameters of a 2-breather realized in an atomic BEC. This object is a 2-soliton, a `nonlinear superposition' of two simple NLSE solitons. The macroscopic parameters we considered were the relative displacement, relative velocity, and relative phase of the two constituent solitons, as well as the difference in the theor numbers of particles. 

A breather can be created in a self-attractive gas by first creating a simple soliton, and then quenching the coupling constant by a factor of four (for details, see Sec.~\ref{sec:concrete_setting}). Within the mean-field theory, immediately after the quench, the spatial symmetry of the setting shows that the constituent solitons of the breather must have strictly zero relative velocity $v$ and displacement $b$. However, the underlying quantum many-body theory predicts that it is only the quantum expectation values of $v$ and $b$ that are zero, while the variances will be non-zero. The principal result of \cite{marchukov2020} was a computation of these variances. In any particular experimental realization, one may therefore be able to observe a separation of the constituent solitons. If the experiment is free enough of other sources of noise, the nonzero values of $v$ or $b$ will be clear manifestations of quantum fluctuations in the soliton---despite the fact that a soliton is a macroscopic object, which, moreover, is in a regime where otherwise the mean-field theory holds to an excellent approximation. 

The approach taken in \cite{marchukov2020} relied on intermediate calculations of so-called quasi-inner products. That method, however, is computationally expensive, and in several cases the calculations could only be completed numerically. The principal contribution of the present work is to simplify and clarify the part of this calculation that concerns the classical fields.

More precisely, here we present a canonical formalism for a class of problems that share the following requirements.

\begin{enumerate}[label*=\arabic*.,topsep=0.5\baselineskip,partopsep=0pt,itemsep=0pt,leftmargin=0em,rightmargin=0em,after=\vspace{0.25\baselineskip},itemindent=2em]
\item To a good approximation, the system of interest should be described by one or more classical, real- or complex-valued fields defined in the continuum or on a lattice, satisfying deterministic equations of motion. Let's call this the `basic model' of the system. 
%
\item The basic model should have a standard Hamiltonian structure. This is as opposed to, for example, the nonstandard Hamiltonian structure of the KdV equation; see \cite{Nutku1984_2007} and Eqs.~(3.15) and (3.16) on p.\ 308 of \cite{takhtajan_book2007}. On the other hand, the NLSE is a good example of a theory with a standard Hamiltonian structure \cite{takhtajan_book2007}. 
\item There should be an extension of the basic model that adds a small stochastic noise to the classical field, so $\psi(x,\,t)$ is replaced by $\psi(x,\,t) + \delta \psi(x,\,t)$.

Usually, this noise originates in some more fine-grained degrees of freedom, which the basic model neglects and which could be either classical or quantum in origin; correspondingly, $\delta \psi(x,\,t)$ is either a classical or a quantum field.

In the case when $\delta \psi(x,\,t)$ is a quantum field, we must have that (i)~the underlying quantum many-body system is describable by a \emph{bosonic} quantum field, and (ii)~the system should be in a weakly interacting regime.

As an example illustrating our method, we will consider the 1D nonlinear Schr{\"o}dinger equation as the basic model. The noise origin will be quantum, so $\delta \psi$ will be a quantum field. But in principle, the noise origin could be classical, such as from either thermal or shot noise. In a thermal case \cite{Malomed1993_R5}, we would need to be able to neglect damping, as might happen when the times of interest are short compared to the typical relaxation time.
\item One should be able to compute correlators of the form $\expectationvalue{\delta \psi(x,\,t)\, \delta \psi(y,\,t)}$, where one or both of the $\delta \psi$'s in the correlator might be $\delta \psi^{*}$ if the field is complex-valued, or $\delta \psi^{\dagger}$ if it is quantum. It is through these correlators that we connect the noise to the classical field model.

In the quantum case, the expectation value should be evaluated with respect to an appropriate state. The choice of this state determines the noise model that is used. In the classical case, the expectation value is a stochastic average, and here, too, the outcome will depend on the choice of a noise model.

We should emphasize that the main thrust of our contribution in the present paper is not about how to choose or construct the correct noise model or, once chosen, how to compute the required correlators. (Although, we will show how to streamline some of the required computations in the particular examples of application of our method that we present below.) Of course, in any given application of our method, one does need to choose a noise model and to compute these correlators. However, as far as the method we are presenting, it is merely assumed that the correlators can be computed \emph{somehow}; they are \emph{inputs} to our method. 

%
\item There should be a canonical system of coordinates that includes, perhaps in addition to continuous (or lattice) variables, also a certain finite number of discrete variables. For lack of a better term, we'll say that these correspond to `macroscopic' degrees of freedom.

The motivating example for this requirement is solitons, which are (`macroscopic') field configurations that exist in a continuum (or on a lattice), but each of which is characterized by just a few canonical variables. These canonical variables typically have simple relationships to the soliton's basic properties such as the initial position and the initial velocity. 

Typically, in a system in which solitons can appear, it is possible for them to exist alongside a non-solitonic, delocalized kind of field dynamics, which is often called `radiation' (see, e.g., Sec.~1.7 of \cite{Ablowitz1981}). The characterization of radiation generally requires not just a few canonical coordinates, but continuum-many (or lattice-many), i.e., at least one canonical pair of \emph{functions}. The argument of these functions is called the `spectral parameter' (see pp.~23 and 232 of \cite{takhtajan_book2007}). This is a continuous parameter (or, for lattice systems, a lattice-like one; see e.g.\ p.~137 of \cite{Ablowitz1981}). 
\item The principal interest should be in computing these `macroscopic' degrees of freedom. The motivating example is again solitons; in systems supporting solitons, we are usually interested in them, i.e., their position, motion, size, etc., and not the radiation (see e.g. the abstract of \cite{Enns1985_632}).
\item In particular, we should be interested in computing the noise-induced fluctuations in the `macroscopic' degrees of freedom. In the case of solitons, this would again be their position, velocity, size, etc.
\end{enumerate}

Provided these requirements are satisfied, our method can, in principle, be used. However, only if the basic model is integrable will it generally be possible to proceed analytically; otherwise, the method must be completely numerical, and in fact it may not be practical. For more on this, see the paragraph following \eqnm{bigvar}.

Section~\ref{sec:fluctuations} presents the principal result of this work: an efficient method of computing the fluctuations in the macroscopic parameters due to the microscopic fluctuations described by $\delta \psi$. In the case of solitons, the macroscopic parameters are the soliton position, velocity, size, etc. 

The final results are simple, but presented rather abstractly, and at that point it might not be clear how they are actually used. Thus, the remainder of the paper is dedicated to their application to a particular problem, namely, solitons in BECs.

In Sec.~\ref{sec:concrete_setting}, we describe this setting in more detail and introduce the problem that can be solved using the results of Sec.~\ref{sec:fluctuations}. This problem is to compute the fluctuation in the parameters of the constituent solitons of a 2-soliton or a 3-soliton upon its creation via a quench in the coupling constant of the NLSE. These fluctuations are quantum in origin; they arise because a BEC is a quantum many-body system, of which the NLSE is a mean-field description. 

The actual computations are presented in Sec.~\ref{sec:computations_of_fluctuations}. Finally, Sec.~\ref{sec:conclusion} presents our conclusions.

The problem treated in Secs.~\ref{sec:concrete_setting} and \ref{sec:computations_of_fluctuations} was first introduced and solved (for a 2-soliton breather) in \cite{marchukov2020}, and we confirm that our results agree with those reported there.

We use the same noise models as that work (if we didn't, we would not have expected our results to match), which we call the `white' noise and the `colored' (or, `correlated') noise. These are explained in Sec.~\ref{sec:computations_of_fluctuations}. Once those are calculated, one needs to compute the effects this microscopic noise has on the solitons' macroscopic parameters. The main difference between the present paper and \cite{marchukov2020} is in how this purely classical step in the computation is done. In principle, the methods of \cite{marchukov2020} should be able to do everything the methods of the present paper do. However, we claim that our method is a significant improvement over the methods of that paper in at least three ways.

The first way is practical. Any method of computation sooner or later runs up against limitations of what can be done with the available computer power, and for the method of \cite{marchukov2020} this happens much sooner than with the method of the present paper. True, in \cite{marchukov2020}, for the white noise model, one was able to evaluate all the required integrals analytically. For the colored noise, however, most integrals had to be done numerically, with some of them taking five days to evaluate. And this was for the 2-soliton case only. To do the 3-soliton, colored case in an acceptable time frame might require a supercomputer, and it is unclear if even that would be enough. In contrast, using the methods of the present paper, even in the case of a 3-soliton with colored noise, we were able to compute the required integrals analytically in a few hours on a laptop.

The second way is conceptual. Reference \cite{marchukov2020} notes in the text after its Eq.~(4) that the relationships between the four pairs of soliton parameters \emph{resemble canonical conjugation (up to constant factors).} Further, in the Supplemental Material (SM) for that paper, we find the following statement in the text following Eq.~(S-17d): 
\begin{quote}
 As was mentioned in the main text the parameter pairs resemble the canonically conjugated variables but the connection is not trivial (see Sec. VI of this Supplemental Material). Moreover, it is tempting to conjecture a connection between the coefficients $C_{\chi\xi}$ and the Poisson brackets of canonical variables and this connection is a subject of future research.
\end{quote}
Finally, in the last section of the SM, the authors tried to uncover as much canonical structure as possible in their formalism, succeeding only partially. The present paper is the fruit of the `future research' mentioned in the quote. Here, the canonical structure takes the central role. Thus, the present paper presents a real conceptual advance, clarifying the role of the canonical structure that was only hinted at, but anticipated and sought after, in \cite{marchukov2020}.

The third way in which our method is an improvement over that of \cite{marchukov2020} is that it contains no unproven assumptions, however reasonable they may be. Namely, in \cite{marchukov2020}, we have the following remark:
\begin{quote}
 In this work, we assume orthogonality of the continuum
fluctuations $\hat{\psi}_\text{cont}$ to the discrete-expansion modes, leaving a rigorous proof of this fact for subsequent work. 
\end{quote}
In the present paper, there is no analogous unproven assumption. This is because \emph{all} degrees of freedom, including the continuum fluctuations, are described by canonical variables; see the paragraph preceding \eqnm{Qvariance}, and especially the footnote [105].

As we keep emphasizing, the principal contribution of this paper is the streamlined classical part of computation. Having said that, we have also improved the evaluation of the field fluctuations due to correlated vacuum in two respects: (i) Following \cite{Castin2009,Gardiner1997_1414,Castin1998,Sinatra2007_033616}, we use the particle-number-conserving Bogoliubov modes, which include additional correction terms compared to those obtained in the standard $U(1)$-symmetry-breaking approach. Surprisingly, it turned out that most final results are not affected by this, which we discuss further in Appendix \ref{sec:AppendixCorrTerms}. (ii) We have streamlined the computation of these correlated-noise fluctuations.

\section{Canonical treatment of relationships between fluctuations}
\label{sec:fluctuations}

We now turn to the presentation of our principal results. A key role will be played by `old' and `new' sets of canonical coordinates, which are connected by a canonical transformation. The terminology of `old' and `new' coordinates (as in, before and after the canonical transformation) is taken from the standard textbook \cite{Goldstein3rd}, as is the convention that, when discussing an abstract canonical transformation, the `old' coordinates are denoted by lower-case letters ($q_j$, $p_j$), and the new coordinates by upper-case ones ($Q_j$, $P_j$). While reading this section, it may be helpful to keep in the back of one's mind the future application of the method to the NLSE. In that context, the `old' canonical coordinates will be the real and the imaginary part of the classical field $\Psi(x,\,t)$, \eqnm{canonicalpair}. The `new' coordinates will be the soliton parameters, four for each constituent soliton, which describe the soliton's norm and initial position, initial velocity, and initial phase. These `new' coordinates are described in detail in Sec.~\ref{sec:math_struct_breathers}. 

Note the remarkable reduction in the number of coordinates: continuum-many for the `old' ones (which are fields), and just eight `new' ones in the case of the 2-soliton (four for a simple soliton). This happens because we consider cases where there is no `radiation' mentioned several paragraphs above. If there were also radiation, then, in addition to the eight solitonic coordinates (each of which is just a real number), we would also require a pair of real-valued \emph{functions}, $\rho(\lambda)$ and $\varphi(\lambda)$, describing the radiation, where $\lambda$ is the so-called spectral parameter (see p.~232 of \cite{takhtajan_book2007}). So radiation requires continuum-many coordinates to characterize it.

A key point is that in the absence of radiation, there are very explicit expressions available expressing the NLSE field in terms of the solitonic parameters. For an $n$-soliton, these expressions are obtained by solving the \emph{linear} $n$-by-$n$ system of equations, see Eqs.~\eqref{nsoliton_eqn}-\eqref{fieldsum}. Looking at that system of equations, we see that the resulting expressions for $\psi(x,\,t)$ will have the form of some algebraic combination of the exponentials of combinations of $x$, $t$, and the soliton parameters. (Some of the parameters also enter algebraically on their own, outside of exponentials.) This is, all things considered, surprisingly simple, given that it describes the solution of a nonlinear partial differential equation. As a consequence, taking derivatives of $\psi$ with respect to the soliton parameters is elementary, if tedious (but these days one just does it in \emph{Mathematica} anyway). 

Now consider the reverse situation, where we are asked to express the soliton parameters as functions (or, rather, functionals) of the field, and then differentiate them with respect to the field. This would seem to be \emph{much} less straightforward. Indeed, in order to even extract the values of the soliton parameters from the wavefunction, one must formulate and solve the appropriate inverse scattering problem \cite{Ablowitz1981,takhtajan_book2007}. In short, in our problem of interest, a direct computation of `$\partial\,\text{old}/\partial \,\text{new}$' is relatively easy, whereas a direct computation of `$\partial\,\text{new}/\partial\,\text{old}$' is very difficult. And as it happens, in order to solve our problem, it is the latter, difficult derivatives that we need. It is at this point that the magic of canonical transformations enters, to which we now turn.

As we said, in our system of interest, the `old' variables are fields. A standard strategy for deriving equations valid for a continuous system is to first obtain the corresponding equations for a related discrete system and then take the continuum limit; see, e.g., Ch.~13 of \cite{Goldstein3rd} or Sec.~2.1 of \cite{Sterman1993}. We will adopt this approach as well.

\subsection{Discrete coordinates}

\newcommand{\oldq}{q}
\newcommand{\oldp}{p}
\newcommand{\newq}{Q}
\newcommand{\newp}{P}

\subsubsection{Quantities of interest: fluctuations}
\label{sec:fluctuations_discrete_basic_form}

Let $\{q_{i}$, $p_{i}\}_{i=1}^{N}$ and $\{Q_{i}$, $P_{i}\}_{i=1}^{N}$ be two sets of canonical coordinates and their conjugate momenta, related by a canonical transformation. As is customary, we will sometimes refer to these as the `old' and `new' sets, or denote them by $(q,\, p)$ and $(Q,\, P)$, respectively. For simplicity, we will assume that all canonical transformations are restricted, i.e., without explicit time dependence, although this restriction is not essential.

For any quantity $x$, let $\delta x$ denote any kind of small change in $x$. In the applications presented later in this paper, these changes will come from quantum fluctuations, but, for now, they could also come from classical stochastic fluctuations or any other kind of perturbation. We write $x=x_0+\delta x$, where $x_0$ is the unperturbed value of $x$. 

In the context of the Bogoliubov decomposition (see Sec.~\ref{sec:vacuum_fluctuations_gen}), the perturbation will be a quantum operator, $\delta \hat x$. We will also assume this operator to be `small', in the sense that in any expression, only the lowest-order terms in $\delta \hat x$ need be retained.

We will be interested in the fluctuations in the `new' variables. Expanding to first order in the fluctuations of the `old' variables yields:
\begin{equation}
\begin{aligned}
\delta Q_{i} &= \sum_{j=1}^{N} \left( \pdv{Q_i}{q_j}\delta q_j+\pdv{Q_i}{p_j}\delta p_j\right),
\\
\delta P_{i} &= \sum_{j=1}^{N} \left( \pdv{P_i}{q_j}\delta q_j+\pdv{P_i}{p_j}\delta p_j\right).
\label{eq:starting_fluct}
\end{aligned}
\end{equation}
In the partial derivatives above, each $Q_i$ and $P_i$ is regarded as a function of all the `old' variables $(q,\, p)$, as given by the canonical transformation. Accordingly, each partial derivative is itself a function of the variables $(q,\, p)$, and is to be evaluated at the unperturbed values $(q_j)_0$ and $(p_j)_0$ for all $j=1,\,\ldots,\,N$. Here $q_j=(q_j)_0+\delta q_j$ and $p_j=(p_j)_0+\delta p_j$.

As previously stated, in our case, what we explicitly know is the expressions for the variables $(q,\, p)$ in terms of the variables $(Q,\, P)$. Therefore, we need to `invert' the partial derivatives in \eqnm{eq:starting_fluct}, so that the derivatives of $Q_i$ and $P_i$ with respect to $q_j$ and $p_j$ are expressed in terms of the derivatives of $q_j$ and $p_j$ with respect to $Q_i$ and $P_i$.

For a general change of variables, this can always be done at least in principle by finding the inverse of the Jacobian matrix of the transformation. However, for large values of $N$, the resulting expressions are generally unwieldy: usually, the best one can do is to write each `inverted' partial derivative as the ratio of a cofactor of the Jacobian matrix and the Jacobian determinant. Fortunately, the fact that our transformation is canonical simplifies matters dramatically, as we will now recall.

\subsubsection{Direct conditions for a canonical transformation}
\label{sec:direct_conditions}

As before, suppose we are given two sets of canonical coordinates and their conjugate momenta, $(q,\, p)$ and $(Q,\, P)$, related by a canonical transformation. Then we have the following `direct conditions' for all $i,\,j=1,\,\ldots,\,N$ (see Eq.~(9.48) in \cite{Goldstein3rd}):
\newcommand*{\pdvt}[2]{\partial #1/\partial #2}
\begin{equation}
\begin{aligned}
\pdv{\newq_i}{\oldq_j}&=\pdv{\oldp_j}{\newp_i} & \pdv{\newq_i}{\oldp_j}&=-\pdv{\oldq_j}{\newp_i} \\
\pdv{\newp_i}{\oldq_j}&=-\pdv{\oldp_j}{\newq_i} & \pdv{\newp_i}{\oldp_j}&=\pdv{\oldq_j}{\newq_i}\,.
\end{aligned}
\label{eq:direct_conditions}
\end{equation}
Here, in $\pdvt{\newq_i}{\oldq_j}$, the variable $\newq_i$ is considered a function of all the variables $(q,\, p)$, as given by the canonical transformation. Similarly, in $\pdvt{\oldp_j}{\newp_i}$, the variable $\oldp_j$ is considered a function of all the variables $(Q,\, P)$, as given by the inverse canonical transformation. Corresponding remarks apply to the other three relations
%
%
\footnote{Although the direct conditions are derived in \cite{Goldstein3rd} under the assumption of a restricted canonical transformation (i.e., one without explicit time dependence), these conditions in fact hold for any canonical transformation, restricted or not. The reason is that they are merely a rewriting of the symplectic condition for a canonical transformation, and that condition is necessary and sufficient for any transformation to be canonical, restricted or not. 
See Derivation 12 in Ch. 9 of \cite{Goldstein3rd}, where the $\bm{M}^{-1}$ on the right-hand side should be $\bm{\widetilde{M}}^{-1}$.}. For our purposes, the relations in \eqnm{eq:direct_conditions} will serve as the required `inversions' of the partial derivatives.

\subsubsection{Lagrange brackets}
\label{sec:Lagrange_brackets}

Because we will be employing the `inverted' partial derivatives, it is the Lagrange brackets, rather than the Poisson brackets, that will naturally appear in our calculations (for example, in the derivations of Eqs.~\eqref{covarsQQ} and \eqref{covars}). We now briefly review their definition, following Sec.\ 9.5 of \cite{Goldstein3rd}. Let $\{\oldq_{i},\, \oldp_{i}\}_{i=1}^{N}$, or $(q,\,p)$ for short, be a set of canonical coordinates and their conjugate momenta. Let $\{y_i\}_{i=1}^{2N}$ be a set of $2N$ independent functions of the variables $(q,\,p)$. We assume that these functions define an invertible change of variables, $y_i=y_i(\oldq_1,\,\ldots,\, \oldq_N,\, \oldp_1,\, \ldots,\, \oldp_N)$. This change of variables is not assumed to be canonical (though it may be). The inverse transformation expresses each $\oldq_{i}$ and $\oldp_{i} $ as a function of $y_1,\, \ldots,\, y_{2N}$. To simplify notation, let us introduce $u$ and $v$ as two of the $y$-variables, e.g., $u=y_\alpha$ and $v=y_\beta$. Then the Lagrange bracket of $u$ and $v$ with respect to the variables $(q,\,p)$ is defined as
\newcommand*{\LagBr}[2]{\{#1,\, #2\}}
\begin{equation}
 \LagBr{u}{v}_{(q,\,p)} = \sum_{k=1}^{N}\left(\pdv{q_k}{u} \pdv{p_k}{v} - \pdv{p_k}{u}\pdv{q_k}{v}\right)\,.
 \label{eq:def_Lagr_brack}
\end{equation}
Like the Poisson bracket, the Lagrange bracket is a canonical invariant. Thus, the subscript $(q,\,p)$ can be omitted, and we can write $\{u,\,v\}$ without ambiguity.

Next, suppose that the two functions $y_\alpha$ and $y_\beta$ (which define $u$ and $v$) are trivial, so that each of $u$ and $v$ is simply one of the original $(q,\,p)$ variables. There are three possibilities: both $u$ and $v$ can be $q$'s, or both can be $p$'s, or one can be a $q$ and the other a $p$. Plugging each case into \eqnm{eq:def_Lagr_brack}, we obtain the fundamental Lagrange brackets:
\begin{equation}
\begin{aligned}
&\LagBr{\oldq_i}{\oldq_j}=\LagBr{\oldp_i}{\oldp_j}=0, \\
&\LagBr{\oldq_i}{\oldp_j}=\delta_{ij}\,.
\end{aligned}
\label{eq:fund_Lagr_brack}
\end{equation}
As we have seen, the properties of the Lagrange bracket in many respects parallel those of the Poisson bracket. However, the Lagrange bracket does not satisfy the Jacobi identity. Instead, it satisfies a different relation (which we will not use here); see Derivation 19 in Ch.~9 of \cite{Goldstein3rd}.

\subsubsection{Fluctuations in terms of inverted partial derivatives}
\label{sec:fluctuations_discrete_inverted_partial_derivatives}

Applying the `direct conditions' of \eqnm{eq:direct_conditions} to \eqnm{eq:starting_fluct}, the latter becomes
\begin{equation}
\begin{aligned}
 \delta Q_{i} &= - \sum_{j=1}^{N} \left(\pdv{q_j}{P_i}\delta p_j- \pdv{p_j}{P_i}\delta q_j\right),
\\
\delta P_{i} &= \sum_{j=1}^{N} \left( \pdv{q_j}{Q_i}\delta p_j - \pdv{p_j}{Q_i}\delta q_j\right).
\end{aligned}
\label{eq:fluct}
\end{equation} 
This is the form that will enable us to evaluate the fluctuations $\delta Q_{i}$ and $\delta P_{i}$.

As an illustration of the properties of the Lagrange brackets, let us use them to verify the correctness of \eqnm{eq:fluct}. For example, to verify the first relation in \eqnm{eq:fluct}, we start from its right-hand side, and substitute the following first-order expansions:
\zeroskips
\vspace{\oldabovedisplayskip}
\begin{align*}
\delta q_j&=\sum_{k=1}^{N}\left( \pdv{q_j}{Q_k}\delta Q_k+\pdv{q_j}{P_k}\delta P_k\right)\\
 \intertext{and}
\delta p_j&=\sum_{k=1}^{N}\left( \pdv{p_j}{Q_k}\delta Q_k+\pdv{p_j}{P_k}\delta P_k\right).
\end{align*}\par
\vspace{\oldbelowdisplayskip}
\restoreskips
\noindent In the double sum that appears, we first exchange the order of summation. Then, for each $k$, we group the terms with a common $\delta Q_k$, which we factor out. We do the same for the terms with a common $\delta P_k$. We get 
\begin{equation*}
\begin{split}
\sum_{k=1}^{N}\sum_{j=1}^{N} \left[ \left(\pdv{q_j}{Q_k}\pdv{p_j}{P_i}-\pdv{p_j}{Q_k}\pdv{q_j}{P_i}\right)\delta Q_k\right. \hspace{5em} & \\
\left.+\left(\pdv{q_j}{P_k}\pdv{p_j}{P_i} - \pdv{p_j}{P_k}\pdv{q_j}{P_i}\right)\delta P_k\right] \hspace{3em} & \\
 & \hspace{-20em} =\sum_{k=1}^{N} \left[\LagBr{Q_k}{P_i}\delta Q_k+\LagBr{P_k}{P_i}\delta P_k\right] \\
 & \hspace{-20em} =\sum_{k=1}^{N} \left[\delta_{ki}\delta Q_k+0\,\delta P_k\right] \\
 & \hspace{-20em} = \delta Q_i,
 \end{split}
\end{equation*}
where in the next-to-last step we used the properties of the fundamental Lagrange brackets, \eqnm{eq:fund_Lagr_brack}. The second relation in \eqnm{eq:fluct} can be verified the same way.

\subsection{Fields}
\label{sec:fields}
In our field-theoretic system of interest, the discrete set of `old' canonical variables $\{q_{i}$, $p_{i}\}_{i=1}^{N}$ is replaced by `old' canonical fields: a one-dimensional coordinate field $q(x)$ and its conjugate momentum field $p(x)$.

The `new' canonical variables, on the other hand, appear in a hybrid form. First, there is a pair of continuous functions of a certain spectral parameter, which are a `new' canonical coordinate and its `new' conjugate momentum. These will not be important in what follows, so we need not discuss them further. Second, in addition to these continuous `new' variables, there is also a certain number of discrete pairs of `new' coordinates and their conjugate momenta. Our main interest will lie in some of these discrete `new' variables.

The most straightforward way to handle the continuous degrees of freedom is to first discretize space onto a regular one-dimensional lattice with spacing $\Delta x$, reproduce the relevant relationships for the discrete variables $\tilde{q}_{i}\approx q(x)/\Delta x,\,\tilde{p}_{i} \approx p(x)/\Delta x$, and finally replace any spatial sums by integrals.

Applying this procedure to \eqnm{eq:fluct}, the fluctuations for the `new' discrete coordinates and momenta can be written as 
\begin{equation}
\begin{aligned}
 \delta Q_{i} &= - \int_{-\infty}^{\infty} \left(\pdv{q(x)}{P_i}\delta p(x)- \pdv{p(x)}{P_i}\delta q(x)\right)\,dx,
\\
\delta P_{i} &= \int_{-\infty}^{\infty} \left( \pdv{q(x)}{Q_i}\delta p(x) - \pdv{p(x)}{Q_i}\delta q(x)\right)\,dx.
\end{aligned}
\label{eq:fluct_cont}
\end{equation}
Here, each of the fields $q(x)$ and $p(x)$ is, in general, a function of all the `new' variables, both discrete and continuous. We will not need to handle (and therefore will not explain) the partial derivatives with respect to the `new' continuous variables.

\subsection{Vacuum fluctuations around a one-dimensional Bose condensate}
\label{sec:vacuum_fluctuations_gen}

In requirement 3.\ listed in the Introduction, we said that the fluctuations may be either classical or quantum in origin. However, in the remainder of this paper we will only be considering quantum fluctuations. They occur because our classical field is an approximate description of a many-body quantum mechanical system.

The underlying quantum model can be written in either the first- or the second-quantized form. For present purposes, the second-quantized form is more important, because its connection to the classical NLSE is more manifest. (Having said that, in Sec~\ref{subsec:underlying_quantum_many-body_model} we will briefly mention the first-quantized form as well.)

\newcommand{\FockVacKet}{\ket{0_\text{F}}}
\newcommand{\FockVacBra}{\bra{0_\text{F}}}
Let us recall the essential ingredients of any 1D, nonrelativistic, bosonic, spin-0 quantum field theory (QFT). The basic objects are the quantum field $\hat{\Psi}(x,\,t)$ and the Fock vacuum $\FockVacKet$. The Fock vacuum is assumed to be the unique (up to overall phase) quantum state of unit norm that is annihilated by $\hat{\Psi}$ for all $x$ and $t$:
\begin{equation}
 \hat{\Psi}(x,\,t)\FockVacKet=0,
 \label{eq:Fock_vac}
\end{equation}
so that the field $\hat{\Psi}$ is also called the annihilation operator field. Its Hermitian conjugate $\hat{\Psi}^\dagger$ is called the creation operator. The two fields are postulated to satisfy the standard canonical equal-time bosonic commutation relations:
\begin{equation}
 \begin{aligned}
 & \left[\hat{\Psi}(x,\,t),\, \hat{\Psi}^\dagger(y,t)\right]=\delta(x-y)\\
 & \left[\hat{\Psi}(x,\,t),\, \hat{\Psi}(y,t)\right]=\left[\hat{\Psi}^\dagger(x,\,t),\, \hat{\Psi}^\dagger(y,t)\right]=0.
 \end{aligned}
 \label{eq:NLSE_field_comm_rel}
\end{equation}

For most of the remainder of this subsection, we will assume that everything happens at a fixed moment of time, and omit explicitly showing the argument $t$. 

The operator $\widehat{N}=\int_{-\infty}^{\infty}\hat{\Psi}^\dagger(x)\hat{\Psi}(x)\,dx$ is called the particle-number operator. Any finite eigenvalue of $\widehat{N}$ must be a non-negative integer (for these and other basic QFT facts reviewed here, see Sec.~A.3 of \cite{Huang1987a}).

\newcommand{\wpsi}{\altChi}

Let $\ket{\wpsi_N}$ be a unit-norm eigenstate of $\widehat{N}$ with eigenvalue $N$. It has the following representation:
\begin{multline}
 \ket{\wpsi_N}=\frac{1}{\sqrt{N!}}\int_{-\infty}^{\infty}\cdots\int_{-\infty}^{\infty}dx_1\cdots dx_N\, \wpsi_N(x_1,\, \ldots,\,x_N) \\
 \times \hat{\Psi}^\dagger(x_1)\cdots\hat{\Psi}^\dagger(x_N)\ket{0}.
 \label{eq:alternative_rep}
\end{multline}
Here $\wpsi_N(x_1,\,\ldots,\,x_N)$ has the interpretation as the first-quantized many-body wavefunction for the system. Because all the field operators in \eqnm{eq:alternative_rep} commute and inner products must be consistent between the first and second quantization,\footnote{The necessity of the second condition is almost never mentioned in the literature, but the fact is that from $\int\cdots\int \,dx_1\cdots dx_N\, f(x_1,\, \ldots,\,x_N) \\
 \times \hat{\Psi}^\dagger(x_1)\cdots\hat{\Psi}^\dagger(x_N)\ket{0}=\int\cdots\int\, dx_1\cdots dx_N\, g(x_1,\, \ldots,\,x_N) \\
 \times \hat{\Psi}^\dagger(x_1)\cdots\hat{\Psi}^\dagger(x_N)\ket{0}$, one cannot conclude that $f=g$; one can only conclude that $S\,f=S\,g$, where $S$ is the summetrization operator. Indeed, we \emph{could} keep $\wpsi_N(x_1,\, \ldots,\,x_N)$ unsymmetrized if we modified the definition of the first-quantization inner product so that it includes an explicit symmetrization. Symmetrized or not, $\wpsi_N(x_1,\, \ldots,\,x_N)$ will satisfy the first-quantized many-body Sch{\"o}dinger equation.} we have that $\wpsi_N$ is completely symmetric with respect to permutations of its arguments.

Our classical field model describes a Bose-Einstein condensate. At the quantum many-body level, the presence of the condensate means that we can use the Bogoliubov decomposition \cite{Bogolubov1947}, in which a bosonic quantum field operator, $\hat{\Psi}$, is written as the sum of a classical mean-field component $\Psi_0$ (the very field from our classical field theory model), and a `remainder' quantum field $\delta \widehat{\Psi}$ \cite{Castin1998}, which is also called the phonon field operator \cite{Gardiner1997_1414}:
\begin{equation}
 \hat{\Psi} = \Psi_0 + \delta \widehat{\Psi}. 
 \label{eq:Bogl_decomp}
\end{equation}
This approximation is valid when the non-condensed fraction of particles is small, i.e., when the system is in a state $\ket{\wpsi_N}$ in which 
\begin{equation}
 \expectationvalue{\delta \widehat{N}}\ll N\,,
 \label{eq:small_non-condensed_fraction}
\end{equation}
 where $\expectationvalue{\delta \widehat{N}} = \int_{-\infty}^{\infty}\expectationvalue{\delta \widehat{\Psi}^{\dagger}(x)\delta \widehat{\Psi}(x)}\,dx$ and the expectation value $\expectationvalue{\cdots}$ is taken with respect to the state $\ket{\wpsi_N}$ \cite{Castin1998,Gardiner1997_1414}.

As we said, we will be interested in the fluctuations---which originate from the presence of the quantum field $\delta \widehat{\Psi}$---of the macroscopic variables characterizing localized states of the one-dimensional Bose condensates. It will turn out that these macroscopic variables can be expressed in terms of the `new' discrete canonical variables.

A natural choice for `old' canonical coordinate and momentum fields is
\begin{equation}
\begin{aligned}
q(x) &= \sqrt{2\hbar}\Re[\Psi(x,\,t)]\\
p(x) &= \sqrt{2\hbar}\Im[\Psi(x,\,t)],
\end{aligned}
\label{canonicalpair}
\end{equation}
The corresponding definitions for the small fluctuations are
\begin{equation}
\begin{aligned}
 \delta \hat q(x) & = \sqrt{2 \hbar} \,\frac{\delta \hat{\Psi}(x)+\delta\hat{\Psi}^\dagger(x)}{2} \\
 \delta \hat p(x) & = \sqrt{2 \hbar}\, \frac{\delta \hat{\Psi}(x)-\delta\hat{\Psi}^\dagger(x)}{2 i} \, . 
\end{aligned}
 \label{lildeltas}
\end{equation}

Now, as far as quantum fluctuations are concerned, it turns out we already know how to compute four basic correlators of the form $\bra{0}\delta \hat u(x)\delta\hat v(y)\ket{0}$, for certain kinds of vacuum states $\ket{0}$, where each of $\delta \hat u(x)$ and $\delta \hat v(x)$ is either $ \delta \hat q(x)$ or $ \delta \hat p(x)$.

To take advantage of this, we will be computing quantities such as $\bra{0}\delta \hat Q_i^2\ket{0}$, which we will be denoting by $\expectationvalue{\delta \hat Q_i^2}$ when it is clear which vacuum state is meant. In that example, we use the first relation in \eqnm{eq:fluct_cont} twice, obtaining
\begin{multline}
  \expectationvalue{\delta \hat Q_i^2} =\\
  \int_{-\infty}^{+\infty} \int_{-\infty}^{+\infty}\Biggl( \pdv{p(x^\prime)}{P_i}\pdv{p(y^\prime)}{P_i} \bra{0}\delta \hat q(x^\prime)\delta \hat q(y^\prime)\ket{0} \\
 -\pdv{p(x^\prime)}{P_i}\pdv{q(y^\prime)}{P_i} \bra{0}\delta \hat q(x^\prime)\delta \hat p(y^\prime)\ket{0}\\
  -\pdv{q(x^\prime)}{P_i}\pdv{p(y^\prime)}{P_i} \bra{0}\delta \hat p(x^\prime)\delta \hat q(y^\prime)\ket{0}\\
 +\pdv{q(x^\prime)}{P_i}\pdv{q(y^\prime)}{P_i} \bra{0}\delta \hat p(x^\prime)\delta \hat p(y^\prime)\ket{0} \Biggr) \dd x^\prime \dd y^\prime.
 \label{bigvar}
\end{multline}

It will turn out that such integrals can be evaluated analytically, in closed form. This lucky circumstance is traceable to two factors: (i)~Due to the integrability of the NLSE, it is possible to write analytically tractable expressions for the `old' coordinates (which are the real and imaginary parts of the NLSE field) in terms of the `new' coordinates (which are the soliton parameters). (ii)~Because the `old' and `new' coordinates are related by a canonical transformation, the derivatives of the `new' coordinates with respect to the `old' ones, which is what we need but which would be very hard to evaluate directly, can be expressed in a simple way (via the `direct conditions' of \eqnm{eq:direct_conditions}) in terms of the derivatives of the `old' coordinates with respect to the `new' ones. And the latter are tractable, thanks to (i).

\section{A concrete setting: the nonlinear Schr{\"o}dinger equation}
\label{sec:concrete_setting}

We will now show how to apply the canonical formalism to the $n$-soliton breathers of the nonlinear Schr{\"o}dinger equation (NLSE). 

\subsection{The underlying quantum many-body model}
\label{subsec:underlying_quantum_many-body_model}

Now we specialize to a system whose Hamiltonian is given by
\begin{multline}
 \widehat{H}=\int_{-\infty}^{\infty}\left(\frac{\hbar^2}{2m} \pdv{\hat{\Psi}^\dagger(x)}{x}\pdv{\hat{\Psi}(x)}{x} \right.\\
 \left.+ \frac{g}{2} \hat{\Psi}^\dagger(x) \hat{\Psi}^\dagger(x)\hat{\Psi}(x)\hat{\Psi}(x)\right)dx,
 \label{eq:QFT_H}
\end{multline}
where $m>0$ and $g$ are constants (see, e.g., Sec.~I.1 of \cite{korepin1997_book}). 

This Hamiltonian gives rise to the following Heisenberg equation of motion for the quantum fields:
\begin{equation}
i \hbar \frac{\partial}{\partial t} \hat{\Psi}(x,\,t) = -\frac{\hbar^2}{2m}\frac{\partial^2}{\partial x^2} \hat{\Psi}(x,\,t) + g \hat{\Psi}^\dagger(x,\,t)\hat{\Psi}^2(x,\,t).
\label{quantum_NLSE}
\end{equation}

This is the quantum nonlinear Schr{\"o}dinger equation. We will return to it shortly.

A state of the form \eqref{eq:alternative_rep} will be an eigenstate of $\widehat{H}$ from \eqnm{eq:QFT_H}, with eigenenergy $E_N$, if and only if the function $\wpsi_N$ is an eigenstate of the Lieb-Liniger Hamiltonian \cite{Lieb1963_1605} $\widehat{H}_N= -\frac{\hbar^2}{2m} \sum_{j=1}^{N}\pdv{^2}{x_{j}^2} + g\,\sum_{j<k}\delta (x_j-x_k)$, with the same eigenenergy.

The Lieb-Liniger model, defined by the first-quantized many-body Hamiltonian $\widehat{H}_N$, describes a system of $N$ identical bosonic particles of spin 0 and mass $m$, whose pairwise interactions are described by the zero-range potential $V(x,y)= g\,\delta(x-y)$. The functions $\wpsi_N$ are the $N$-body time-independent wavefunctions of the quantum-mechanical system. This model is a quantum many-body integrable system. It is exactly solvable through the Bethe ansatz \cite{gaudin1983_book_english,sutherland2004_book}, both in the $g>0$ case \cite{Lieb1963_1605,mcguire1964_622,yang1967_1312,yang1968_3,dorlas_orthogonality_1993,korepin1997_book}, which corresponds to repulsive interactions between the particles, and in the $g<0$ case \cite{Berezin1964_21,mcguire1964_622,BrezinZinnJustin1966,yang1968_3,calogero1975_265,Lai1989b,Calabrese2007_150403,Calabrese2007_P08032,yurovsky2017_220401}, which corresponds to attractive interactions between the particles. The latter is the case that is relevant for us.

In particular, we are interested in the attractive Lieb-Liniger model such that the particles can move anywhere along the whole one-dimensional real line. Let us introduce the center-of-mass (CoM) coordinate
$
R=\frac{1}{N} \sum_{j=1}^{N} x_{j}
$
and the relative coordinates $z_{j}$, $j=1,\,\ldots, N-1$. The latter may be defined in various ways, but for our purposes, the following will do: $z_{j}=x_{j+1}-x_{1}$. \footnote{This particular system of CoM and relative coordinates is not orthogonal, but it is simple to define, and its Jacobian determinant of the change of variables from the Cartesian coordinates is 1. Depending on what one wishes to compute, other choices of relative coordinates may be better, e.g., \cite[Eq.~(94)]{Castin2009} or \cite[Eq.~(5)]{Liu2010_023619}.}

Since the system is translationally invariant, the Hamiltonian commutes with the CoM momentum operator $\hat{P}_{\text{CoM}}=-i\hbar\, \partial/\partial R$; thus the energy eigenvalues can always be chosen so that they are also the eigenstates of $\hat{P}_{\text{CoM}}$. Such eigenstates factorize as $\phi(x_{1},\,\ldots,\, x_{N})=e^{i K R} \varphi_{\text{rel}}(z_{1},\,\ldots,\,z_{N-1})$, where $K$ is the eigenvalue of $\hat{P}_{\text{CoM}}$. Because these states have definite values of the CoM momentum, their CoM position is completely delocalized.

Now, the ground-state $N$-body wavefunction of the attractive Lieb-Liniger model on the real line is given by
\cite{calogero1975_265,Castin2001_419}
\begin{equation}
\phi_{\text{G}}(x_{1},\,\ldots,\, x_{N}) = \lmcal{N} \exp \left(-\frac{m |g|}{2\hbar^{2}}\sum_{1\leqslant j < k \leqslant N} \left|\vtsp x_{j}-x_{k}\right|\right)\,.
 \label{eq:true_quantum_many-body_ground_state}
\end{equation}
Here $\lmcal{N}$ is a normalization factor to be discussed momentarly. Note that $\phi_{\text{G}}$ can be rewritten so it depends only on the relative coordinates. Therefore, the CoM momentum is zero. It follows that the CoM position is completely delocalized. One consequence is that the state is not normalized in the usual sense. There are several ways of dealing with this \cite{calogero1975_265,Castin2001_419}, but here is the one used in \cite{calogero1975_265}: we demand that the state be normalized once the CoM position is fixed (say, to zero), so that the normalization condition becomes
\begin{equation*}
 \int_{-\infty}^{\infty}d x_{1}\cdots\int_{-\infty}^{\infty}d x_{N} \,\delta(R)\left|\phi_{\text{G}}(x_{1},\,\ldots,\, x_{N})\right|^{2}=N\,.
\end{equation*}
The value of $\lmcal{N}$ that follows from this condition is known \cite{calogero1975_265}, but we won't need it for our discussion.

It is informative to consider the 1-body density
\begin{multline*}
 \rho_{N}(x)= \\
 \int_{-\infty}^{\infty}d x_{1}\cdots\int_{-\infty}^{\infty}d x_{N} \,\delta(R)\vtsp\delta(x_{1}-x)\left|\phi_{\text{G}}(x_{1},\,\ldots,\, x_{N})\right|^{2}\,.
\end{multline*}
(Because of the bosonic symmetry of $\phi_{\text{G}}$, no generality is lost in singling out $x_{1}$ in this expression.) While the exact expression is available  \cite{calogero1975_265}, here we are mostly interested in the large-$N$ expansion \cite{calogero1975_265}:
\begin{equation}
 \rho_{N}(x)= \frac{N}{4\xi} \sech^{2}\left(\frac{x}{2\xi}\right)\left[\rule[-2pt]{0pt}{10pt}\smash{1+\mathcal{O}(1/N)}\right]\,,
 \label{eq:quant_soliton_density}
\end{equation}
where
\begin{equation}
\xi=\frac{\hbar^2}{m |g| N} 
\label{eq:soliton_width}
\end{equation}
is a characteristic length. We now see that $\phi_{\text{G}}$ describes the $N$ particles being exponentially localized within a `lump' of size $\xi$, where the position of the lump (which is the CoM coordinate) is completely delocalized. This is the quantum many-body version of the simple (or, elementary) bright soliton. Below, we will see that the classical-field (or, mean-field) description of density (i.e., $\left|\Psi_{\text{simp.\ sol.}}\right|^{2}$, where $\Psi_{\text{simp.\ sol.}}$ is given in \eqnm{eq:simple_soliton}) recovers exactly the leading-order term in \eqnm{eq:quant_soliton_density}.

As we said, the CoM position of the state $\phi_{\text{G}}$ is completely delocalized. In contrast, in actual experiments the solitons are localized (as we will see in a moment). Nevertheless, the Lieb-Liniger model in general and the state $\phi_{\text{G}}$ in particular are experimentally very relevant, as we now discuss.

\subsection{Relevance to experiments}
\label{subsec:relevance_to_experiments}

Though experiments with ultracold gases nowadays use a variety of atomic and molecular species, in our discussion here we will have in mind principally alkali metals from lithium to cesium \cite{metcalf99,Leggett2001_307,Zhai2021}. Here we will present a brief sketch of how the Lieb-Liniger model can be realized in such experiments; for more details, see Refs.~\cite{Leggett2001_307,Bloch2008_885}.

We begin by briefly discussing the experimental realization of 3D BECs, after which we describe the additional trapping that makes such systems effectively one-dimensional. 

Let us first consider the interaction between two alkali atoms. For a large enough interatomic distance $r$, it is valid to describe the pairwise interaction by an interatomic potential $V(r)$ that, for large $r$, has the van der Waals asymptotic form $-C_{6}/r^6$ \cite{Gribakin1993_546,Leggett2001_307}.

Next, we point out several important length scales. In the relevant experiments, it is crucial that some of them are much greater than others, as we shall see. These lengths are: 1.\ the thermal de Broglie wavelength of the atoms $\lambda_{\text{dB}}=\sqrt{\frac{2\pi\hbar^2}{m k_{\text{B}}T}}$, where $k_{\text{B}}$ is the Boltzmann constant and $T$ is the temperature; 2.\ the typical interatomic distance $r_{\text{at}}=1/n_{\text{dens}}^{1/3}$, where $n_{\text{dens}}$ is the number density of the gas; and 3.\ the range of interaction $b_{\text{int}}$. Under ordinary circumstances, $b_{\text{int}}$ is the characteristic van der Waals length $r_{\text{vdW}}$, defined as $r_{\text{vdW}}=(2 m_{\text{r}} C_{6}/\hbar^{2})^{1/4}$, where $m_{\text{r}}$ is the reduced mass. We will see below that in the experiments most relevant to us, the use of the magnetically tuned Feshbach resonance \cite{Chin2010_1225} makes $b_{\text{int}}$ substantially smaller than $r_{\text{vdW}}$. But it will turn out that this only makes all the required inequalities hold even better. 

In BEC experiments, the length scales we just introduced compare as follows: 1.\ The ultracold gases are dilute, in the sense that $b_{\text{int}} \ll r_{\text{at}}$. This means that (i) it is legitimate to think of the pairwise atomic interactions in terms of scattering; and (ii) the probability of three or more atoms simultaneously coming close enough to interact is negligible, so we only need to consider two-body processes. 2.\ A BEC, being a quantum degenerate gas, is characterized by the condition $r_{\text{at}} \leqslant \lambda_{\text{dB}}$. 3.\ From the first two inequalities we get $b_{\text{int}} \ll \lambda_{\text{dB}}$. But the de Broglie wavelength is related to the typical wavenumber through $k=2\pi/\lambda_{\text{dB}}$, meaning that $k\,b_{\text{int}} \ll 1$. This means that the system is in the regime of low-energy scattering. Furthermore, since the scattering potential decays faster than $1/r^3$, we conclude that (i) only the $s$-wave scattering is important, and (ii) the scattering properties are completely determined by a single parameter, the $s$-wave scattering length $a_{\text{sc}}$ (see \cite{Pera2023_90} and \S~``The scattering of slow particles'' in \cite{Landau1991quantum}). 

Under these conditions, the range of interaction $b_{\text{int}}$ is arguably better quantified by $|a_{\text{sc}}|$ than by $r_{\text{vdW}}$. Since these two are normally of the same order, this consideration often does not matter. However, many experiments take advantage of a magnetically tuned Feshbach resonance \cite{Chin2010_1225} to set $a_{\text{sc}}$ to essentially any value. In particular, in the experiments most relevant to us (i.e., \cite{Luo2020_183902,Jin2024_dissertation}), $a_{\text{sc}}$ is Feshbach-tuned to values between $-5 a_{0}$ and $-0.1 a_{0}$, where $a_{0}$ is the Bohr radius. These turn out to be, respectively, 13 and 650 times smaller in magnitude than $r_{\text{vdW}}$. As we mentioned above, this just means that the inequalities $b_{\text{int}} \ll r_{\text{at}}$ and $b_{\text{int}} \ll \lambda_{\text{dB}}$ hold even better when $b_{\text{int}}$ is set to $|a_{\text{sc}}|$. For definiteness, here is how the various length scales compare in \cite{Luo2020_183902}, where $a_{\text{sc}}=-0.15 a_{0}$: $|a_{\text{sc}}|/r_{\text{at}} \approx \num{4e-5}$, and $|a_{\text{sc}}|/\lambda_{\text{dB}} \approx \num{4e-6}$. For future reference, we note that here the density was taken to be the maximal density at the peak of the `mother soliton' (see Sec.~\ref{sec:creating_breathers}).

Note that $b_{\text{int}} \ll r_{\text{at}}$ in particular means that the so-called gas parameter, $n_{\text{dens}} b_{\text{int}}^{3}$, is much less than 1, so that the gas is in the weakly interacting regime (see the discussion following Eq.~(360) in \cite{Castin2001_1}). Taken together, all these facts imply that certain well-known approximate descriptions should be valid, namely, the 3D classical field (or, mean-field) model and the Bogoliubov theory. Their 1D counterparts will be introduced in Secs.~\ref{sec:classical_field} and \ref{sec:correlated_noise}, respectively.




Since the atomic interaction is completely determined by the $s$-wave scattering and its $a_{\text{sc}}$, one can replace the true interatomic potential by the simplest potential with the same low-energy scattering properties. Arguably, this will be some zero-range potential, i.e., one proportional to the Dirac delta-function (where one must keep in mind that a $D$-dimensional delta-function potential requires a regularization prescription unless $D=1$ \cite{Wodkiewicz1991_68,Farrell2010_817}). The usual choice is the Huang-Fermi pseudopotential \cite{Huang1987a} characterized by the same scattering length: $\widehat{U}_{\text{HF}}=g_{\text{3D}}\, \delta^{3}(\vec{r})\,\hat{\mathcal{R}}$. Here $\vec{r}$ is the relative coordinate of the two atoms, $g_{\text{3D}}=2\pi\hbar^{2}a_{\text{sc}}/m_{\text{r}}$, and $\hat{\mathcal{R}}$ is the regularization operator, defined by $\hat{\mathcal{R}}\,\psi(r,\,\theta,\,\phi)=\frac{\partial}{\partial r}\left[r \psi(r,\,\theta,\,\phi)\right]$. A wavefunction $\psi(\vec{r})$ will be a solution of the Schr{\"o}dinger equation with $\widehat{U}_{\text{HF}}$ as the potential if and only if $\psi$ satisfies the Wigner-Bethe-Peierls contact condition \cite{Bethe1935_146,Wigner1933_253} $\psi(\vec{r}) = A(1/r-1/a_{\text{sc}})+\mathcal{O}(r)$ for $r\to 0^{+}$, where $A$ is a constant (i.e., independent of $\vec{r}$). The job of the regularization operator is to remove the $A/r$ term, which would otherwise produce a divergence once the delta function sets $\vec{r}=0$ \cite{Dunjko2011_461}. 
For further details and various generalizations, see \cite{Grossmann1984_1742,Wodkiewicz1991_68,Olshanii2001_010402,Idziaszek2006_013201,Kanjilal2006_060701,Pricoupenko2007_2065,Pricoupenko2008_170404,Stampfer2010_052710,Farrell2010_817,Le2019_065203}; for a mathematically rigorous treatment of the pseudopotential, see Ch.~I.1 of \cite{albeverio_solvable_2005}.

The next step toward an experimental realization of the Lieb-Liniger gas is to place the BEC into a highly anisotropic, cigar-shaped harmonic trap. Experimentally, this trap may be optical, magnetic, or a combination of both \cite[Ch.~4]{Jin2024_dissertation}. The trap is characterized by the transverse trapping frequency $\omega_\perp$ and a much smaller longitudinal trapping frequency $\omega_z$. The longitudinal trapping can in principle be taken all the way to zero or even inverted, providing an expulsive potential \cite{Khaykovich2002}. It is important that all relevant energy scales of the many-body state are significantly smaller than the energy-level spacing of the transverse confinement. This results in a system in which the transverse part of the 3D wavefunction is `frozen' in the ground state of the transverse potential. Then the only dynamical degrees of freedom are in the longitudinal direction, making the system effectively one-dimensional. As a practical condition for one-dimensionality, one demands that both $k_{\text{B}} T$ and $|\mu|$ be much smaller than $\hbar \omega_\perp$ \cite{Moritz2003_250402}, where $\mu$ is the chemical potential (which can be negative, as in \eqnm{eq:chemical_potential}).

Once the system is effectively one-dimensional, the scattering of two atoms can be described by a purely 1D Schr{\"o}dinger equation with potential given by \cite{Olshanii1998,Bergeman2003_163201,Dunjko2011_461} $V(z)=g\,\delta(z)$, where $z$ is the relative coordinate and $g=2 \hbar \omega_\perp a_{\text{sc}}$ \footnote{The expression for $g$ given in the text should further be multiplied by $1/(1 - C\,a_{\text{sc}}/a_{\perp})$, where $a_{\perp}=\sqrt{\hbar/(m_{\text{r}}\omega_{\perp})}$ and $C=-\zeta(1/2)=1.46035\ldots$, with $\zeta$ the Riemann zeta function. However, this extra factor is 1 to an excellent approximation in systems that interest us here; e.g., in \cite{Luo2020_183902}, $1$ minus the factor equals $\num{4e-6}$. But there are situations, e.g. \cite{Haller2009_1224}, where this factor can be very large in magnitude and of either sign.}. In experiments in which $a_{\text{sc}}$ can be tuned via the magnetic Feshbach resonance, $g$ can consequently be arranged to have essentially any value, positive or negative. In principle, attractive BECs are subect to collapse \cite{Gammal2001_055602,PhysRevA.97.053604}. However, as long as the 1D condition $|\mu|\ll \hbar \omega_\perp$ holds, the gas will be stable.

As a result of all these reductions, the effectively 1D BEC becomes a realization of the Lieb-Liniger model (or, equivalently, the quantum NLSE) \cite{Yurovsky2008_61}. Indeed, experiments show that this model works very well \cite{%
Moritz2003_250402, 
Paredes2004_277, 
Kinoshita2004_1125, 
Kinoshita2006_900, 
Haller2009_1224, 
Fabbri2015_043617, 
Meinert2017_945, 
Huang2025_eadv3727}. 
For some results that go beyond the approximations introduced above, see, e.g., \cite{Yurovsky2006_163201,Jachymski2017_052703}.

\subsection{The classical field model and its breather solutions}
\label{sec:classical_field}

Given that one has a BEC near its ground state, the Bogoliubov decomposition from \eqnm{eq:Bogl_decomp} works in the regime of weak interactions \cite{Lieb1963_1605,Castin2001_419,Castin2004_89}. Let us give a heuristic argument that the Lieb-Liniger soliton in \eqnm{eq:true_quantum_many-body_ground_state} is, perhaps somewhat counterintuitively, in the regime of weak interactions as long as it consists of many particles.

Let us consider for a moment the repulsive, $g>0$ Lieb-Liniger model. For this system, the thermodynamic limit is well-defined ($N,\,L\to \infty$ with the 1D number density $n_{1D}=N/L$ kept constant; here $L$ is the system size and $N$ is the number of particles). In this case, the interaction strength is quantified by the dimensionless parameter $\gamma_{\text{LL}}=m g/(\hbar^{2} n_{\text{1D}})$. Note that, for a fixed coupling constant, the interaction strength decreases with increasing number density; so the denser the gas, the closer to the non-interacting limit it is, unlike in 3D \cite{Lieb1963_1605,Petrov2000_3745,Dunjko2001_5413}. As counterintuitive as this is, the fact that $g$ and $n_{\text{1D}}$ must enter as $g/n_{\text{1D}}$ follows already from dimensional analysis.

Now, for the attractive Lieb-Liniger soliton in \eqnm{eq:true_quantum_many-body_ground_state}, the thermodynamic limit is not well defined. (This can be seen in several ways. First, for a fixed density, the ground state energy is proportional to $N^{2}$ rather than $N$ \cite{Lieb1963_1605}, so that the energy is not an extensive quantity. Second, the width $\xi$ given in \eqnm{eq:soliton_width} goes to 0 in this limit.) Nevertheless, we can heuristically argue as follows. Let us estimate an effective number density, $n_{\text{1D,eff}}$, as $N/\xi$. This gives an effective Lieb-Liniger parameter $\gamma_{\text{LL,eff}}=1/N^{2}$. So the condition for weak interactions for the ground state of the attractive Lieb-Liniger gas is simply that $N\gg 1$. A more careful argument, arriving at the same condition, is given in the text surrounding Eq.~(99) of  \cite{Castin2001_419}. The condition is amply satisfied in the experiment of \cite{Luo2020_183902}, where $N\sim \num{5e+4}$. (That the conditon must involve $N$ alone also follows already from dimensional analysis; see Sec.~\ref{sec:dimensional_analysis}.)

Let us then apply the Bogoliubov decomposition from \eqnm{eq:Bogl_decomp} to the quantum NLSE field from Eqs.~\eqref{eq:QFT_H} and \eqref{quantum_NLSE}:
\begin{equation}
 \hat{\Psi}(x,\,t)=\Psi(x,\,t)+\delta \hat{\Psi}(x,\,t).
 \label{eq:NLSE_Bogl_decomp}
\end{equation}

In this subsection we will discuss the classical field $\Psi(x,\,t)$. Note, however, that the residual quantum field $\delta \hat{\Psi}(x,\,t)$ will also be of great interest to us, because it is this field that causes the quantum fluctuations we are interested in computing. We will treat this field in Sec.~\ref{sec:correlated_noise}, using the Bogoliubov theory.

The classical field $\Psi(x,\,t)$, which is also variously referred to as the mean-field approximation and the order parameter, represents the Bose-condensed part of the gas \cite{Leggett2001_307}. It satisfies the (classical) NLSE, which, in this context, is also called the (1D) Gross-Pitaevskii equation:
\begin{equation}
i \hbar \frac{\partial}{\partial t} \Psi = -\frac{\hbar^2}{2 m}\frac{\partial^2}{\partial x^2} \Psi
+ g \left|\Psi\right|^2\Psi\,.
\label{classical_NLSE}
\end{equation}
The corresponding Hamiltonian is
\begin{equation}
  H = \int_{-\infty}^{+\infty} \frac{\hbar^2}{2m}\left(\abs{\pdv{x} \Psi} ^2 + \frac{g}{2} |\Psi|^4 \right) \dd x\,.
  \label{classical_Hamiltonian}
\end{equation}

We have mentioned that the underlying quantum many-body system, the Lieb-Liniger model, is an integrable system. The classical NLSE is also an integrable system, exactly solvable using the inverse scattering method \cite{Zakharov1971_118, Ablowitz1981,Novikov1984,takhtajan_book2007}.

Because the NLSE is Galilean invariant, if $\Psi(x,\,t)$ is a solution, then it remains a solution upon a displacement by $x_{0}$ and a boost by a velocity $v$: $\Psi(x-x_{0}-vt,\,t)\,e^{\frac{i}{\hbar}\left[m v (x-x_{0})-\frac{1}{2}m v^{2} t\right]}$.

The norm $N = \int_{-\infty}^{+\infty} |\Psi(x,\,t)|^2 \, dx$ is the classical counterpart of the particle-number operator $\widehat{N}$. Just as $\widehat{N}$ is a conserved quantity in the underlying quantum many-body case (corresponding to the total number of particles in the system), so the norm $N$ is a conserved quantity in the classical field theory. It corresponds to the number of particles in the BEC, which, due to \eqnm{eq:small_non-condensed_fraction}, is very nearly the same as the total number of particles in the system. The total mass is then given by $m N$. Because of these simple relationships among the total mass, the total number of particles, and the norm, it is customary in the literature to refer to them interchangeably. In the case when $\Psi(x,\,t)$ is composed of several constituent solitons, we may likewise refer interchangeably to the mass, number of particles, and norm of each constituent soliton.

When $g<0$, we say that we have an attractive, or focusing, case of the NLSE. This case supports bright soliton solutions \cite{Zakharov1971_118,Satsuma1974_284,takhtajan_book2007,Prilepsky2007}. The ground state is the elementary (or, simple) soliton of norm $N$, given by
\begin{equation}
 \Psi_{\text{simp.\ sol.}}(x,\,t)=\frac{1}{2}\sqrt{\frac{N}{\xi}}\,\sech\left(\frac{x-x_{0}}{2\xi}\right)\,e^{-\frac{i}{\hbar}\mu\,t+i\Theta}\,.
 \label{eq:simple_soliton}
\end{equation}
Here $x_{0}$ is the position of the soliton peak (and CoM), while $\Theta$ is a constant global phase, both of which can have any value; $\xi$ is the soliton width, which is the same as in the quantum many-body case, \eqnm{eq:soliton_width}; and 
\begin{equation}
\mu=-\frac{m g^{2} N^{2}}{8\,\hbar^{2}}
 \label{eq:chemical_potential}
\end{equation}
is the chemical potential of the simple soliton. Note again that $\left|\Psi_{\text{simp.\ sol.}}\right|^{2}$ recovers exactly the leading-order term in \eqnm{eq:quant_soliton_density}.

The attractive NLSE further supports multisoliton solutions. These are `nonlinear superpositions' of the simple solitons (perhaps displaced and boosted), in a sense explained in Sec.~\ref{sec:math_struct_breathers}. Of special interest to us will be multisolitons whose constituent simple solitons have the same positions and velocities, but different norms. These solutions are called `breathers', because their density profiles oscillate in time with a well-defined period, essentially due to the fact that their chemical potentials are different; see \cite[Eq.~(5)]{Prilepsky2007}.

\subsection{Creating breathers}
\label{sec:creating_breathers}

Consider two NLSEs, 1 and 2, whose coupling constants are $g_1$ and $g_2$, respectively. Suppose $g_1=g_2/n^{2}$, with $n$ an integer. It turns out that there is a relation between (i) the elementary soliton solution of NLSE~1 and (ii) the so-called \emph{odd-norm-ratio} (ONR) breather solution of NLSE~2. The latter is an $n$-soliton breather such that the norms of its constituent solitons are in the ratio $1:3:5:\cdots:(2n-1)$. Assume that the total norm is the same for both (i) and (ii). Remarkably, the wavefunction of (ii) at the moment when its density profile is the widest is identical to the wavefunction of (i) \cite{Satsuma1974_284}. This makes it possible to experimentally realize ONR breathers, by first creating an elementary soliton and then using the Feshbach resonance to quench the coupling constant by a factor of $n^2$ \cite{Carr2002,Di_Carli2019_123602,Luo2020_183902}. In this context, the initially created elementary soliton (i) is called the `mother soliton', while the constituent solitons of (ii) are called the `daughter' solitons.

Following the quench, the experiment \cite{Luo2020_183902} has $|\mu|\sim \hbar \omega_\perp$, which is expressed as $N/N_{\text{c}}=1$ in the paper, where $N_{\text{c}}$ is the critical number of atoms beyond which collapse occurs (assuming other parameters are kept fixed). Thus, (i) the one-dimensionality is marginal and (ii) in principle, the system is subject to collapse. However, because the breather is not the ground state of the condensate, it is metastable \cite{PhysRevA.97.053604}, and no collapse occurs on the time scales of the experiment.

In a breather, the relative velocities and positions of the constituent solitons are zero. That this must be so is also clear from the quench protocol that creates the breather. After all, the elementary soliton is completely symmetric about its center of mass, and both NLSE 1 and NLSE 2 respect this symmetry. Thus, the post-quench field $\Psi$ must have the same symmetry. But if any relative velocities or displacements of the constituent solitons were non-zero, that symmetry would be violated; thus, these relative velocities and displacements must be zero. Consequently, any observed post-quench nonzero relative velocity or displacement of the daughter solitons must be due to beyond-mean-field effects. And if the experiment is sufficiently clean and controlled, then the dominant beyond-mean-field effect will be quantum many-body fluctuations, to be discussed in Secs.~\ref{sec:white_noise} and \ref{sec:correlated_noise}.

These beyond-mean-field effects can in principle also be computed directly from the Lieb-Liniger model, due to its integrability. This was done in \cite{ yurovsky2017_220401}. However, it was necessary to use a computer algebra system, and due to the loss of numerical significance calculations became unreliable for $N>23$. When the numerically significant results were used to extrapolate to large $N$, the results agree with the correlated-noise · results of \cite{marchukov2020}---and of the present paper---to 20\%.

\subsection{Quantum state of the mother soliton in experiments}
\label{subsec:quantum_state_mother_soliton}

Let us now concentrate in particular on the realization of the bright soliton from \eqnm{eq:true_quantum_many-body_ground_state}, which serves as the `mother soliton' in \cite{Luo2020_183902}. Recall that the true ground state of the attractive Lieb-Liniger model, i.e., the quantum bright soliton given in \eqnm{eq:true_quantum_many-body_ground_state}, is translationally invariant. However, the presence of the longitudinal harmonic trap breaks the translational symmetry, and the experimentally realized mother soliton has a center of mass (CoM) that is at least somewhat localized near the bottom of the longitudinal trap. Indeed, if one measures the density profile of the mother soliton, it will be well-described by the mean-field description $|\Psi_{\text{simp.\ sol.}}|^{2}$ from \eqnm{eq:simple_soliton}, with $x_{0}=0$. The relationship between the states given by Eqs.~(\ref{eq:true_quantum_many-body_ground_state}) and (\ref{eq:simple_soliton}) were studied in \cite{calogero1975_265,Castin2001_419}.

To get some insight into the quantum many-body nature of the mother soliton, we follow the discussion in \cite{Castin2009}. First, in the presence of a harmonic trap (and this property fails if the trap is not harmonic), the CoM and relative degrees of freedom are separable, just as they are if there is no trap. Furthermore, the energy characterizing the internal degrees of freedom, which is the magnitude of the chemical potential $\mu$ from \eqnm{eq:chemical_potential}, is much greater than the energy characterizing the longitudinal trap, $\hbar \omega_{z}$. Equivalently, the soliton width $\xi$ is much smaller than the width of the ground state of the trap, $\sqrt{\hbar/(m\omega_z)}$. This means that the relative degrees of freedom of the soliton will be largely unaffected by the presence of the trap. Now recall that  $\phi_{\text{G}}$ in \eqnm{eq:true_quantum_many-body_ground_state}, which is the ground state of the system \emph{without} the longitudinal harmonic trap present, actually depends only on the relative degrees of freedom. It follows that the wavefunction of the mother soliton (\emph{with} the longitudinal harmonic trap present) can be written to an excellent approximation as the product $\Phi(R)\,\phi_{\text{G}}(x_{1},\,\ldots,\, x_{N})$, where $\Phi(R)$ is the wavefunction of the CoM. In fact, one would expect this form to hold approximately even in a non-harmonic trap, as long as the typical size of the trap is much greater than~$\xi$.

In the ideal case, $\Phi(R)$ should be the ground state of the trap. In practice, however, during the preparation stage of current experiments, one is never able to `cool' the CoM motion all the way to the ground state. So there is always some uncertainty as to what is the true $\Phi(R)$, which at any rate depends nontrivially on the details of the experimental preparation. For example, $\Phi(R)$ is also subject to time-dependent phase-diffusion effects \cite{Lewenstein1996,Castin1998}. Further, if the longitudinal trapping is turned off, it will be subject to further spreading due to dispersion. In short, the most we can say about $\Phi(R)$ is that it is at least somewhat localized to the center of the trap, and otherwise it is largely unknown.

In Sec.~4 of \cite{Castin2009}, it was investigated whether the act of opening the longitudinal trap introduces significant internal excitations in the gas. The answer is that, under typical experimental conditions, it does not. In more detail: it turns out that, in the large-$N$ limit, to the leading order in $N$, the mean number of internal excitations $N_{\text{exc}}$ can be computed within the classical field model (the condition for the validity of this treatment is that $N\gg |\mu|/(\hbar \omega_{z})$). The conclusion was that already for $|\mu|/(\hbar \omega_{z})\sim 10$, we have that $N_{\text{exc}}/N$ is as low as $10^{-5}$ even when the trap is opened suddenly. As one would expect, $N_{\text{exc}}/N$ gets smaller both as $|\mu|/(\hbar \omega_{z})$ gets larger and as the trap opening becomes more gradual.

In short, as long as one is interested in properties that depend only on the internal, relative degrees of freedom of the soliton (which includes, e.g., the relative displacement $b$ and relative velocity $v$ of the constituent solitons of a breather), one can assume that the quantum state of the mother soliton is the one given in \eqnm{eq:true_quantum_many-body_ground_state}.

\subsection{The mathematical structure of breather solutions}
\label{sec:math_struct_breathers}

Our approach is predicated on the assumption that soliton parameters are canonical coordinates. This wasn't the case in the original treatment of the NLSE \cite{Zakharov1971_118}, or in many of the subsequent treatments, such as \cite{Gordon1983}. However, Faddeev and Takhtajan provided the required canonical treatment in their book \cite{takhtajan_book2007}.

In Chapter 3 \S 7 of Part~I of that book, one finds a reversible canonical transformation for the NLSE based on the so-called inverse scattering data. Since these results are of critical importance for our project, we will dedicate this section to reviewing them.

Faddeev and Takhtajan study the NLSE in the form
\begin{equation}
  i \pdv{\psi}{t} = - \pdv[2]{\psi}{x} + 2 \kappa \abs{\psi}^2 \psi \,,
  \label{TakEq}
\end{equation} 
with the norm $\int_{-\infty}^{+\infty} |\psi(x,\,t)|^2 \, dx=N$. 
We will only be considering the case $\kappa<0$. The connection to the full SI version in \eqnm{classical_NLSE} is explained in Appendix~\ref{sec:AppendixUnits}, but it can be briefly summarized by saying that we are working in a system of units in which $\hbar=1$, $m=1/2$, and $g=2\kappa$, with $\kappa$ dimensionless (corresponding to Option 1 in Sec.~\ref{Faddeev-Takhtajan_form} of Appendix~\ref{sec:AppendixUnits}). 

Let $\{g\}_\text{SI}$ be the numerical value of $g$ in the SI system. The sign of $\kappa$ must match the sign of $\{g\}_\text{SI}$, but the magnitude of $\kappa$ can be arbitrary: for a fixed $\{g\}_\text{SI}$, a change in the magnitude of $\kappa$ results in a change of the units of length and time. This is a bit counterintuitive, but it becomes less so once one notes the following scaling property of solutions of \eqnm{TakEq}: if $\psi(x,\,t)$ is a solution of that equation, then it can be written as 
\begin{equation}
\psi(x\,,\,t)=\sqrt{\kappa}\,\varphi(\kappa x,\,\kappa^2 t)\,,
 \label{scaling_property}
\end{equation}
where $\varphi(x,\,t)$ is a solution of \eqnm{TakEq} with $|\kappa|=1$ (where both $\psi$ and $\varphi$ have the same norm, $N$). Indeed, for our purposes, we could right away set $\kappa=-1$ without loss of generality; for more on that, see Appendix~\ref{sec:AppendixUnits}. We will not do so just yet, so that the expressions we list appear exactly as in Faddeev and Takhtajan. 

A generic solution of the NLSE is, in a sense which will become clearer below, a `nonlinear superposition' of a finite number of constituent simple solitons plus a non-solitonic component. The latter is also called the \emph{radiation} component, because it spreads---radiates---quickly, becoming more and more spatially diluted. In terms of the discussion in Sec.~\ref{sec:fields}, the constituent solitons give rise to the `new' discrete canonical variables, while the non-solitonic component gives the two `new' continuous variables. When we talk about $n$-soliton solutions, we mean purely solitonic solutions, without any non-solitonic component.
In Faddeev and Takhtajan's formulation of an $n$-soliton solution, each constituent soliton is characterized by four real-valued parameters, which form two canonical pairs. (They prove that these parameters are indeed related to the `old' variables of \eqnm{canonicalpair} via a canonical transformation.) For the $j$th constituent soliton, the members of the first canonical pair are denoted $q_j$ and $p_j$, and of the second, $\rho_j$ and $\phi_j$. (The parameters $q_j$ and $p_j$ are not to be confused with the fields $q(x,\,t)$ and $p(x,\,t)$.) Here $q_j$ and $p_j$ are proportional, respectively, to the initial position and the velocity of the $j$th constituent soliton, while $\rho_j$ and $\phi_j$ are the norm (i.e., the number of particles) and the phase of the same constituent soliton $j$. The actual initial ($t=0$) position and velocity of the $j$th soliton are given, in `Gordon units' (see the text preceding \eqnm{GordonToF-T}), by $x_j=2 q_j/\rho_j$ and $v_j=p_j/2$. There is a slight caveat as far as the interpretation of $x_j$, which however is not important for our purposes; see Appendix~E of \cite{dunjko2015_150100075}, especially the final three paragraphs. The total norm $N$ (i.e., the total number of particles) of an $n$-soliton is given by the sum of all the $\rho_j$'s: $N=\sum_{j=1}^{n} \rho_j$.

Breather solitons are the $n$-solitons in which all the constituent solitons have the same positions and the same velocities, but no two have the same norm \cite{Zakharov1971_118}. 

Let us introduce the so-called `scattering data' $\lambda_j$ and $\gamma_j$, which will become important shortly:
\begin{equation}
\begin{aligned}
\lambda_j &=-\frac{\kappa }{2} \left(p_j+i \rho _j\right) \\
\gamma_j &=\exp \left(q_j-i \phi _j\right) \exp \left(-i t \lambda _j^2\right) \exp \left(i x \lambda _j\right) \,.
\end{aligned}
\label{scatteringdata}
\end{equation}

Note that for each $j$, we have that $\lambda_j$ and $\gamma_j$ are associated only with the $j$th constituent soliton. Moreover, the $x$ that appears in $\gamma_j$ is the $x$ in \eqnm{TakEq} (and not the `initial' position $x_j$ introduced above).

According to Chapter 3 \S 5 of Part~I of \cite{takhtajan_book2007}, multi-soliton solutions for this system are then found by first solving the following linear system for the auxiliary fields $\psi_k(x,\,t)$:
\begin{equation}
  \sum_{k=1}^n B_{jk} \psi_k = 1
  \label{nsoliton_eqn}
\end{equation}
Here $n$ is the number of constituent solitons, and
\begin{equation}
  B_{jk}(x,\,t)=\frac{\gamma^*_k +\gamma_j^{-1}}{\lambda^*_k-\lambda _j},
  \label{Mjk}
\end{equation}
with $^*$ indicating complex conjugation.

Note that the individual auxiliary fields $\psi_k(x,\,t)$ have no simple interpretation. In particular, it is not the case that each is associated with a different constituent soliton, since each depends on all the $\lambda_j$ and $\gamma_j$, for all $j$.

The full $n$-soliton solution is then constructed as 
\begin{equation}
\psi(x,\,t) = \frac{1}{\sqrt{|\kappa|}} \sum_k^n \psi_k(x,\,t).
\label{fieldsum}
\end{equation}

When we say that an $n$-soliton is a `nonlinear superposition' of its constituent solitons, we mean that each constituent soliton is characterized by its own parameters $\lambda_j$ and $\gamma_j$, which enter nonlinearly into the solution \eqref{fieldsum}.

Similar results, albeit without a canonical structure, are found in \cite{Zakharov1971_118} and \cite{Gordon1983}.

It will be useful for later to note that solutions with the same norm ratios of constituent solitons but different total norms (remember that $N=\sum_j\rho_j$) are related by simple rescalings. Although we are not aware of any reference that states this, one can see directly from Eqs.~(\ref{nsoliton_eqn})-(\ref{fieldsum}) that for all $\xi$,
\begin{equation}
 \psi(x,\,t\,|\, \rho_j,\,p_j) = \xi\,\psi(\xi x,\,\xi^2 t\,|\,\rho_j/\xi,\,p_j/\xi)\,.
 \label{lambda_rescalings}
\end{equation}
This equation is written in a bit of a shorthand: the last two arguments stand for the $2n$ values ($j=1,\,\ldots,\,n$) of the canonical momenta characterizing the $n$-soliton; the corresponding $2n$ canonical coordinates are the same on both sides and are not shown.


In what follows, we will require $n=2$ and $n=3$ soliton solutions with arbitrary parameters for each constituent soliton (because we will need to take derivatives with respect to these parameters). Due to the complexity of these solutions, we will not display them. However, they are easily generated by software such as \textit{Mathematica}, by solving the system \eqref{nsoliton_eqn} and applying \eqnm{fieldsum}.

\subsection{The problem to be solved}

As we mentioned, the ONR $n$-soliton breathers can be realized by first creating an elementary soliton (the `mother' soliton), which may be undergoing substantial center-of-mass motion (i.e., it might be `hot'). Next, we quench the coupling constant by a factor of $n^2$. This produces an ONR $n$-soliton breather (whose constituent solitons we will call `daughter' solitons), created at the point of its breathing cycle when its width is maximal. The actual field $\Psi(x,\,t)$ does not change at the moment of quench; only its subsequent evolution does. Since the daughter solitons constitute a breather, their relative velocities are exactly zero, even though the center-of-mass motion of the breather might be substantial. 

Any nonzero relative velocities are a signature of the presence of fluctuations in the mother soliton. The present-day experiments are so well controlled that the dominant source of fluctuations is the intrinsic quantum fluctuations of the mother soliton. Thus, any observed post-quench separation of the constituent solitons would be a beyond-mean-field, quantum-many-body effect. 

In addition to nonzero relative velocities, the quantum fluctuations of the mother soliton also cause the post-quench constituent solitons to be created already slightly displaced from each other, and to have norms whose ratios deviate from the ratios of consecutive odd integers. The latter implies that the field will have a small non-solitonic component \cite{Satsuma1974_284}, which, however, will quickly radiate away (i.e., it will quickly reach the edges of the physical trap and be mostly lost from the system). Finally, the relative phases of the constituent solitons will also deviate from their classical-field predictions.

The problem we want to solve is to compute the immediate post-quench variances of the distributions of the constituent solitons' relative velocities and relative positions around their classical-field zero values (which are zero). We also want to compute the variance of relative phases and norms. Any non-zero values for these variances are due to the quantum fluctuations of the matter field $\Psi$ at the moment of quench.

The basic principle that makes this calculation possible is this: although the quench changes the Hamiltonian instantaneously and discontinuously (well, very quickly and sharply, at any rate), the post-quench change in the wavefunction is much slower and smoother. This is because that change is due solely to the regular time evolution under the post-quench Hamiltonian; the fact that the new Hamiltonian was arrived at through a sudden change in one of its parameters does not matter. This is true at both the classical and the quantum-many-body levels.

Thus, the following procedure is valid: in an expression such as \eqnm{bigvar}, the quantum fluctuation---i.e., the factors such as $\bra{0}\delta \hat q(x^\prime)\delta \hat p(y^\prime)\ket{0}$---are computed for the mother soliton. (We will compute these fluctuations for two different noise models; for a naive `white noise' model, the computation is straightforward, while for the more realistic `colored noise' model, we will use known results from the literature.)
On the other hand, the partial derivatives such as $\pdv{q(x^\prime)}{P_i}$ are computed assuming $q(x^\prime)$ corresponds, via \eqnm{canonicalpair}, to the breather $\Psi$. Again, as a function of $x$, the field $\Psi$ just before and just after the quench is the same, and so the `old' canonical variables \eqref{canonicalpair} are the same. However, the `new' canonical variables just before the quench are different from those just after the quench. 

In more detail: before the quench, there is only a single soliton and thus $q(x)$ and $p(x)$ are functions of (i.e., parameterized by) only $4$ discrete `new' variables: $q_1$, $p_1$, $\rho_1$, and $\phi_1$. This parameterization is the one in Eq.~(S-27) of the Supplemental Material for \cite{marchukov2020}, where our $q_1$, $p_1$, $\rho_1$, and $\phi_1$ correspond to (and are simple functions of) the parameters $B$, $V$, $N$, and $\Theta$ in that equation. 

Immediately after the quench, even though $q(x)$ and $p(x)$ are unchanged as functions of $x$, they are now functions of (i.e., parameterized by) $4n$ discrete `new' variables: $q_j$, $p_j$, $\rho_j$, and $\phi_j$ for $j=1,\,\ldots,\, n$. For example, the parameterization of the $2$-soliton solution is the one in Eq.~(S-3) of the same Supplemental Material. In that equation, the 8 parameters are the center-of-mass (CoM) parameters $B$, $V$, $N$, and $\Theta$, and the relative parameters $b$, $v$, $n$, and $\theta$. (Here $n$ is not the number of constituent solitons; rather $n=N_1-N_2$, where $N_1$ and $N_2$ are the norms of the constituent solitons.)

\newcommand{\unitLG}{L_\text{G}}
\newcommand{\unitTG}{T_\text{G}}
\newcommand{\unitVG}{V_\text{G}}

It will be helpful to keep in mind the physical interpretation of these parameters. To do that, it will be helpful to introduce the `Gordon parameters' \cite{Gordon1983}. For each constituent soliton $j$, these are its norm $N_j$, initial position $x_j$, initial velocity $v_j$, and the initial phase $\theta_j$. Of these, we have already mentioned $x_j$ and $v_j$; they are the initial position and velocity when expressed in terms of the `Gordon units' for length and velocity (see Appendix~\ref{sec:AppendixUnits}). The relevant Gordon units are these:
\begin{align}
 \unitLG&=\frac{\hbar^2}{m|g|} & \unitTG&=\frac{\hbar^3}{mg^2} & \unitVG&=\frac{\unitLG}{\unitTG}=\frac{|g|}{\hbar}\,.
 \label{Gordon_units}
\end{align}
The norm and the phase do not change as we change units. The connection with the Faddeev-Takhtajan parameters is as follows:
\begin{equation}
\begin{aligned}
N_j&=\rho_j & \theta_j&=-\phi_j+q_j p_j/\rho_j\\
x_j&=2 q_j/\rho_j & v_j&=p_j/2 \,.
\end{aligned}
\label{GordonToF-T}
\end{equation}
Most of the CoM and relative parameters are related in an obvious way to the Gordon parameters:
\begin{equation}
\begin{aligned}
 N &=N_1+N_2 & n&=N_2-N_1\\
 B&=\frac{N_1x_1+N_2x_2}{N_1+N_2} & b&=x_2-x_1\\
V&=\frac{N_1v_1+N_2v_2}{N_1+N_2} & v&=v_2-v_1\,.
\end{aligned}
\end{equation}
In particular, $B$ and $b$ are the CoM and relative positions, and $V$ and $v$ are the CoM and relative velocities, expressed in the Gordon units, \eqnm{Gordon_units}.
The links between phases are a bit more complicated; we won't show them (but see Eq.~(S-36) in the Supplemental material for \cite{marchukov2020}) because we use these links only as intermediaries for deriving the links beteen the CoM/relative coordinates and the Gordon parameters. That result is given in Table~\ref{table:COMCanonical}. The reason we need these latter links is that, as mentioned in the Introduction, for the $2$-soliton NLSE breather, the desired variances and covariances have already been computed in \cite{marchukov2020}, and we want to make sure those results match ours.


\section{The computation of fluctuations}
\label{sec:computations_of_fluctuations}

The causes of the fluctuations in the post-quench constituent solitons' relative velocities and other quantities of interest are the quantum fluctuations of the field $\hat \Psi$. They enter our formalism through the quantum expectation values such as $\bra{0}\delta \hat q(x^\prime)\delta \hat p(y^\prime)\ket{0}$. To evaluate these factors, we need a model for the quantum fluctuations, which comes down to a choice for the `vacuum' state $\ket{0}$ with respect to which those quantum expectation values are to be taken.

As in \cite{marchukov2020}, we will be considering two different models, i.e., two different `vacuum' states $\ket{0}$. The first one is a simple white-noise model. The other is a more realistic model with a particular kind of correlated noise.

At this point, we set $\kappa=-1$ in \eqnm{TakEq}; as explained above, this can be done without loss of generality.

\subsection{White-noise vacuum}
\label{sec:white_noise}
\newcommand{\WhiteVacKet}{\ket{0_\text{W}}}
\newcommand{\WhiteVacBra}{\bra{0_\text{W}}}

The white-noise model of quantum noise is implemented as follows. We apply Eqs.~\eqref{lildeltas}, \eqref{eq:delta_psi_commutation_rels}, and \eqref{eq:whitenoise_vac} to \eqnm{bigvar}. The resulting expressions contain the quantum expectation values $\bra{0}\delta \hat q(x^\prime)\delta \hat p(y^\prime)\ket{0}$, etc. In the white-noise model, we take that the state $\ket{0}$ is the unit-norm state that is annihilated by $\delta \hat{\Psi}(x,\,t)$. We denote it by $\WhiteVacKet$:
\begin{equation}
 \begin{aligned}
  \delta \hat{\Psi}(x)\WhiteVacKet &= 0 \\
  \WhiteVacBra\delta\hat{\Psi}^\dagger (x) &= 0.
  \end{aligned}
  \label{eq:whitenoise_vac}
\end{equation}

It should be obvious that $\WhiteVacKet$ is different from the Fock vacuum state $\FockVacKet$. After all, from Eqs.~\eqref{eq:NLSE_Bogl_decomp} and \eqref{eq:Fock_vac}, it follows that if any Bose-condensed atoms are present, then $\delta \hat{\Psi}(x,\,t)$ cannot annihilate the Fock vacuum $\FockVacKet$.

Although both $\FockVacKet$ and $\WhiteVacKet$ are referred to as `vacua', they correspond to vastly different physical situations. The Fock vacuum corresponds to there being no actual particles present, i.e., no atoms. In contrast, the white-noise vacuum corresponds to an absence of ``bare'' fluctuations (i.e., fluctuations created by $\delta \hat{\Psi}$).

Next, consider Eqs.~\eqref{eq:NLSE_field_comm_rel} and \eqref{eq:NLSE_Bogl_decomp}. It is evident that the quantum-fluctuation field $\delta \hat{\Psi}(x,\,t)$ also satisfies the canonical equal-time commutation relations,
\begin{equation}
 \begin{aligned}
& \qty[\delta \hat{\Psi}(x,\,t),\,\delta \hat{\Psi}^\dagger(y,t)] = \delta(x-y) \\
& \qty[\delta \hat{\Psi}(x,\,t),\,\ \delta \hat{\Psi}(y,t)] =\qty[\delta \hat{\Psi}^\dagger(x,\,t),\,\ \delta \hat{\Psi}^\dagger(y,t)] =0.
\end{aligned}
\label{eq:delta_psi_commutation_rels}
\end{equation}
This concludes the definition of the `white-noise' model. Together with Eqs.~(\ref{eq:whitenoise_vac}) and (\ref{lildeltas}), it allows us to compute all the required expectation values.

The white-noise model is very commonly used in quantum optics \cite{Haus1996, Haus1990, Lai1993, Yeang1999, Haus2000}, and has been imported into atomic physics as well \cite{Opanchuk2017,marchukov2020}. It is usually described by saying that quantum fluctuations have the form of uncorrelated random noise \cite{marchukov2020}. It is uncorrelated in the sense that the equal-time two-point correlation function (evaluated with respect to the $\WhiteVacKet$ state) vanishes for any nonzero separation of the two points: $\expval{\delta \widehat{\psi}(x,\,0)\,\delta \widehat{\psi}^{\dagger}(y,\,0)} = \delta(x-y)$, which follows immediately from the commutation relations in \eqnm{eq:delta_psi_commutation_rels}, the annihilation condition in \eqnm{eq:whitenoise_vac}, and the fact that the state $\WhiteVacKet$ has unit norm (see also Eq. (31) in \cite{Yeang1999}, where one must remember that for optical solitons discussed there, space and time are reversed relative to the atomic ones; see Eq. (1) in that paper). As for why this is referred to as `white noise', let us reinterpret $\expval{\delta \widehat{\psi}(x,\,0)\,\delta \widehat{\psi}^{\dagger}(x+z,\,0)}$ as the autocorrelation function $R(z)$. Now recall that the power spectral density (i.e., the Fourier transform) of $R(z)=\delta(z)$ is constant, i.e., it contains equal power at all frequencies, i.e., it is `white'. 

If a noise model is to count as different from the white-noise one, it will have to be the case that $\expval{\delta \widehat{\psi}(x,\,0)\,\delta \widehat{\psi}^{\dagger}(y,\,0)}$ (evaluated with respect to the vacuum appropriate for that noise model) does \emph{not} equal $\delta(x-y)$. Such a noise model will thus be `uncorrelated'. The corresponding spectral density will then not contain equal power at all frequencies, so that such a noise model will not be `white'; thus, it will be `colored'. So while there is only one white-noise model, there could in priciple be many colored-noise ones. In Sec.~\ref{sec:correlated_noise}, we will consider a colored-noise model that is physically well-justified in the context of NLSE breathers.
 
The white noise model would be exactly correct if the mother soliton were a Hartree product of noninteracting single-particle wave functions, all having the shape of the mother soliton \cite{marchukov2020}.

In our context, in which the `colored noise' model in Sec.~\ref{sec:correlated_noise} is much better justified physically, one might wonder why we bother with the `white noise' model at all. One reason is that it presents a simplified testbed for computational methods and analytical development, while still yielding results that rarely turn out to differ by more than 50\% from those computed using the more accurate vacuum. Another reason is simply that it is a widespread model with which practitioners have much experience. Had we only used the `colored noise' vacuum, an obvious question would be how well the simpler `white-noise' approximation would do, and whether the additional complexity of the colored-noise model is truly necessary.

Back to computations. In what follows, $q(x')$ and $p(x')$ are the `old' coordinates, \eqnm{canonicalpair}. $Q_j$ stands for either of $q_j$ or $\phi_j$, the canonical coordinates of the $j$th soliton; $P_j$ stands for either of the corresponding canonical momenta, either $p_j$ or $\rho_j$, respectively. For example, in \eqnm{Qvariance}, if $\expval{\delta \hat Q_j^2}=\expval{\delta \hat \phi_2^2}$, then the $P_j$ in the derivatives inside the integral is $\rho_2$.

The partial derivatives are first evaluated using the symbolic expressions for $q(x)$ and $p(x)$ obtained from \eqnm{canonicalpair}, where $\Psi(x,\,t)$ in those equations is set to the $\psi(x,\,t)$ obtained from Eqs.~(\ref{scatteringdata})-(\ref{fieldsum}). Only then are the soliton parameters set to the values given in Table~\ref{table:2sol_initialcond} (and $t$ is set to zero, which is why we don't show it explicitly as an argument of the $q(x)$ and $p(x)$ fields inside the partial derivatives). 

It is true that in the presence of fluctuations, the field $\psi(x,\,t)$ acquires a radiation component. But this component is described by a completely independent degree of freedom (see item 2.\ on p.~126 of \cite{takhtajan_book2007}). Therefore, although formally we should first take the partial derivatives, and only then set the radiation component to zero, the result is the same if one sets the radiation component to zero first \footnote{\label{footnote:independece_argument}Here is a sketch of how this can be shown explicitly from the formalism of Chapter~II of Part~I of \cite{takhtajan_book2007}; all the formulas referenced in this footnote are from that chapter. We begin by taking a partial derivative (with respect to a discrete degree of freedom) of both sides of Eq.~(2.4). The corresponding partial derivatives of $\psi$ enter through Eq.~(2.5); this is, in effect, just an indirect way of computing them. Next, we take into account Eqs.~(2.2), (2.3), and especially (2.18) and (2.28). We note that the discrete degrees of freedom enter only through the function $B(\lambda)$. The radiation itself is described by $b(\lambda)$ in Eq.~(2.13); when it is absent, $b(\lambda)=0$, and all three matrices $G(\lambda)$, $\tilde{G}_{+}(\lambda)$ and $\tilde{G}_{-}(\lambda)$ are just the identity matrix (compare with Eqs.~(5.2) and (5.17)). It is then easy to see that, when taking the partial derivative of Eq.~(2.4), regardless of whether $b(\lambda)$ is set to zero before or after taking the derivative, we end up with the same equation in the end. It follows that the partial derivative of $\psi$ with respect to a discrete degree of freedom is also the same whether the radiation is set to zero before or after differentiation.}.

Using the white-noise model, we find that \eqnm{bigvar} reduces to a single integral,
\begin{equation}
  \expval{\delta \hat Q_j^2} = \frac{1}{2}\int_{-\infty}^{+\infty}\qty[\qty(\pdv{q(x^\prime)}{P_j})^2+\qty(\pdv{p(x^\prime)}{P_j})^2] \dd x^\prime.
  \label{Qvariance}
\end{equation}
Here the prefactor would be $\frac{\hbar}{2}$ (inherited from Eqs.~(\ref{lildeltas})), but $\hbar=1$ in the system of units we are currently using. Secondly, even if we had not set $\kappa$ to $1$, the scaling property of \eqnm{scaling_property} would ensure that $\kappa$ dissapears from these integrals upon the `$u$-substituion' $u=\kappa x$. These remarks apply to all the integrals for the expectation values of the variances and covariances.

The same procedure can be done to find also the initial fluctuations in $P_j$ as well as all required covariances. For completeness, the initial fluctuation in $P$ is
\begin{equation}
   \expval{\delta \hat P_j^2} = \frac{1}{2}\int_{-\infty}^{+\infty}\qty[\qty(\pdv{q(x^\prime)}{Q_j})^2+\qty(\pdv{p(x^\prime)}{Q_j})^2] \dd x^\prime.
  \label{Pvariance}
\end{equation}

It turns out that for covariances, the terms analogous to the middle terms in \eqnm{bigvar} (which in that case have imaginary coefficients) form a Lagrange bracket. Therefore, for covariances one must only calculate
\begin{multline}
  \expval{\delta Q_j \delta Q_k} = \frac{1}{2} \int_{-\infty}^{+\infty} \left(\pdv{q(x^\prime)}{P_j}\pdv{q(x^\prime)}{P_k} \right.\\ 
  \left. +\pdv{p(x^\prime)}{P_j}\pdv{p(x^\prime)}{P_k}\right) \dd x^\prime 
  \label{covarsQQ}
\end{multline}
or
\begin{multline}
  \expval{\delta Q_j \delta P_k} = \frac{i}{2} \{Q_j,P_k\} - \frac{1}{2}\int_{-\infty}^{+\infty} \left(\pdv{q(x^\prime)}{P_k}\pdv{q(x^\prime)}{Q_j} \right. \\
  \left. +\pdv{p(x^\prime)}{P_k}\pdv{p(x^\prime)}{Q_j}\right) \dd x^\prime \,.
  \label{covars}
\end{multline}

It is useful to realize that even if computations above are done with solitons whose total norm is unity (i.e. $\rho_j=\rho^0_j$ with $\sum_j\rho^0_j=1$, one can immediatly write the answer for the case where $\rho_j=N \rho^0_j$, so the total norm is $N$ but all the norm ratios are the same. This is possible because of the relation (\ref{lambda_rescalings}) and the fact that the derivatives are evaluated at $p_j=0$ (see Tables~\ref{table:2sol_initialcond} and \ref{table:3sol_initialcond}). Let us explain this further. 

Note first that $q(x)$ and $p(x)$ that are linear combinations (via \eqnm{canonicalpair} and \eqnm{fieldsum}) of the $\psi_k(x,\,t)$'s and their complex conjugates, where latter are the solutions of the linear system \eqnm{Mjk}. Next, suppose we write the integrals in Eqs.~(\ref{Qvariance})-(\ref{covars}) with $q(x)$ and $p(x)$ that were ultimately obtained from the solutions of the linear system (\ref{Mjk}) with $\rho_j$'s set to to $N \rho^0_j$. These are the integrals we actually want to compute. Now we are going to rewrite them in terms of $q(x)$ and $p(x)$ that were ultimately obtained from the solutions of the linear system (\ref{Mjk}) with $\rho_j$'s set to $\rho^0_j$.

To do this, we apply \eqnm{lambda_rescalings}, in the form of
\[
\psi(x,\,t\,|\,N \rho^0_j,\,p_j) = N\,\psi(N x,\,N^2 t\,|\,\rho^0_j,\,p_j/N)\,,
\]
to all the $\psi_k(x,\,t)$'s and their complex conjugates. We see right away that the prefactor $N$ of $\psi$ translates to an overall factor of $N^2$ in front the integrals, since each term inside the integral has either a product of two $q(x)$'s, or of two $p(x)$'s, or of one of each. Next, since $x'$ now appears as $N x'$, we can do the `$u$-substituion' $u=N x'$. This contributes an overal factor of $1/N$, so the overall prefactor at this point is $N$. Next, every partial derivative with respect to $p_j$ (or $p_k$) contributes another overall factor of $1/N$ due to the chain rule. Finally, every partial derivative with respect to $\rho_j$ (or $\rho_k$) also contributes an another overall factor of $1/N$. This is also due to the chain rule. Let $\bar{\rho}_j=N \rho^0_j$. The derivative we start with is $\partial/\partial\bar{\rho}_j$, but now we require it to be $\partial/\partial\rho^0_j$. According to the chain rule, 
\[\pdv{}{\bar{\rho}_j}=\pdv{\rho^0_j}{\bar{\rho}_j}\pdv{}{\rho^0_j}=\frac{1}{N}\pdv{}{\rho^0_j}\,.
\]
%
%
Putting it all together, the integrals get an overall prefactor of $N^{1-c}$, where $c$ is the number of times that the partial derivatives with respect to $P_j$ (i.e., $\rho_j$ or $p_j$) or $P_k$ (i.e., $\rho_k$ or $p_k$) appear as factors in an individual term inside the integral. Finally, note that although the momenta $p_j$ now enter the $\psi_k$'s as $p_j/N$, the initial values of $p_j$ are always zero. So, once the initial values of the parameters are imposed, the rescaling of the $p_j$'s doesn't change the value of the $\psi_k$'s, but only of their derivatives with respect to $p_j$.

So for example, $c=2$ in \eqnm{Qvariance} (because of the squares) and in \eqnm{covarsQQ}, and so we get an overall prefactor of $1/N$. In \eqnm{Pvariance}, we have $c=0$, and so we get an overall prefactor of $N$. Finally, $c=1$ in the $\expectationvalue{Q_iP_j}$ covariance (\ref{covars}), and so we get no overall prefactors involving $N$.

This matches the $N$-dependences listed in Table~\ref{table:results_2sol_canonical}, which were obtained by using wavefunctions correctly normalized to $N$. However, in the 3-soliton calculations, we set $N=1$, so one should use the prescription we just described to restore the scalings with $N$. Moreover, it will become obvious that the same prescription will work for the colored noise, which is to be expected, since the presence of colored noise should only slightly modify the white-noise values of the purely numerical prefactors in the final results for the variances and covariances. so we set $N=1$ in all computations for that noise.

\subsubsection{2-soliton breather}

We will now directly compute the initial quantum fluctuations of the 2-soliton breather using the formalism we have developed, and show that we can recover the values reported in \cite{marchukov2020} obtained by much more computationally taxing methods. 

The 2-soliton breather is created by a 4-fold quench of the coupling constant $g$. At the initial time, $t=0$, the mother soliton is converted into two daughter solitons with norm ratios of $1:3$. As explained before, the daughter solitons are `born cold', with zero initial relative velocity, although there can be a difference in their phases. The initial values for the daughter soliton parameters as used in our computations are presented in Table \ref{table:2sol_initialcond}.

\begin{table}[h] 
\centering
\begin{tabular}{|c|c|c|c|c|c|c|c|c|}
\hline
 Parameter & $q_1$ & $q_2$ & $p_1$ & $p_2$ & $\rho_1$ & $\rho_2$ & $\phi_1$ & $\phi_2$ \\ \hline
Initial value & $0$ & $0$ & $0$ & $0$ & $\frac{3}{4}N$ & $\frac{1}{4}N$ &0 & $-\pi$\\
 \hline
\end{tabular}
\caption{Initial values of parameters for the 2-soliton breather.}
\label{table:2sol_initialcond}
\end{table}
The basic ingredient in the computation is the solution~\eqref{fieldsum}. As we said before, an explicit form of this solution can be found in Eq.~(S-27) of the Supplemental Material for \cite{marchukov2020} (though the variables have to be changed according to Table~\ref{table:COMCanonical}). However, the solution can be easily found (especially using, e.g., Mathematica) from Eqs.~\eqref{scatteringdata}-\eqref{fieldsum} for $n=2$ with parameters given in Table~\ref{table:2sol_initialcond}.

The relevant partial derivatives of the solution~\eqref{fieldsum}---as they enter into Eqs.~\eqref{Qvariance}, \eqref{Pvariance}, \eqref{covarsQQ}, and \eqref{covars} via \eqnm{canonicalpair}---are taken before evaluating any expressions using the values of the initial parameters. 

While the initial wavefunction is real in this case, the necessary derivatives of $p(x)$ may not be. The results of these calculations are presented in Table \ref{table:results_2sol_canonical}. The integral portions of the expression in \eqref{covars} evaluate to zero for all parameters, so the initial fluctuations are given only by $i/2$ times the Lagrange bracket in question. For example, $\expectationvalue{q_1p_1} = i/2 $, while $\expectationvalue{p_1q_1} = -i/2 $, and $\expectationvalue{q_1p_2} = 0 $ following the properties in \eqref{eq:fund_Lagr_brack}.

\begin{table} 
\centering
\begin{tabular}{|c|c|c|c|c|}
\hline
 \makecell{Expectation\\Value} & \makecell{Initial\\Fluctuation} & Covariance & \makecell{Initial\\Fluctuation} \\
\hline
$\expectationvalue{\delta q_1^2}$ & $\frac{1}{N}\left(\frac{2 \pi ^2}{5}+3\right)$ & $\expval{\delta q_1\delta q_2}$ & $\frac{1}{N}\left(\frac{4 \pi ^2}{15}-1\right)$\\ 
$\expectationvalue{\delta q_2^2}$ & $\frac{1}{N}\left(\frac{2 \pi^2}{5}-1\right)$ & $\expval{\delta p_1\delta p_2}$ & $-\frac{1}{35}N$\\
$\expectationvalue{\delta p_1^2}$ & $\frac{3}{70}N$ & $\expval{\delta\rho_1\delta\rho_2}$ & $\frac{1}{5}N$\\
$ \expectationvalue{\delta p_2^2}$ & $\frac{5}{42}N$ &$\expval{\delta\phi_1\delta\phi_2}$ & $-\frac{1}{N}\left(1+\frac{4 \pi^2}{105}\right)$\\
$\expectationvalue{\delta \rho_1^2}$ & $\frac{3}{10}N$ & $\expectationvalue{Q_iP_j}$ & $ \frac{i }{2}\delta_{ij} $ \\ 
$\expectationvalue{\delta \rho_2^2}$ & $\frac{3}{10}N$ &$\expectationvalue{P_iQ_j}$ & $ -\frac{i }{2}\delta_{ij} $\\
$\expectationvalue{\delta \phi_1^2}$ & $\frac{1}{N}\left(1+ \frac{2 \pi^2}{35}\right)$ &&\\ 
$\expectationvalue{\delta \phi_2^2}$ & $\frac{1}{N}\left(\frac{7}{3} + \frac{10 \pi^2}{63}\right)$ &&\\
 \hline
\end{tabular}
\caption{Initial quantum fluctuations of the two-soliton breather's canonical parameters, obtained for the white-noise vacuum state. 
We have that $\expval{\delta q_1\delta q_2}=\expval{\delta q_2\delta q_1}$, and similarly for $\expval{\delta p_1\delta p_2}$, $\expval{\delta \rho_1\delta \rho_2}$, and $\expval{\delta \phi_1\delta \phi_2}$. Covariances not listed, such as $\expval{\delta \\q_j\delta \phi_k}$, are zero.
}
\label{table:results_2sol_canonical}
\end{table}

Table~\ref{table:results_2sol_canonical} is our principal result for the $2$-soliton with white noise. However, in order to compare our results to those previously obtained in \cite{marchukov2020}, we need to translate these variances to the variances in the CoM and relative coordinates. This is done by applying the standard error propagation formulas to the formulas in Table~\ref{table:COMCanonical}.

\begin{table} 
\centering
\begin{tabular}{|c|c|}
\hline
 \makecell{CoM or rela-\\tive parameter} & Equivalent in canonical coordinates \\ 
 \hline
$N$ & $\rho_1+\rho_2$ \\ 
$V$ & $\frac{p_1 \rho_1 +p_2 \rho_2}{2\qty(\rho_1+\rho_2)}$ \\
$B$ &$\frac{2\qty(q_1+q_2)}{\rho_1+\rho_2}$\\
$\Theta$ & $-\frac{\qty(\rho_1+\rho_2)^2 \qty(\phi_1+\phi_2+\pi)+\qty(p_1-p_2)\qty(q_1+q_2)\qty(\rho_1-\rho_2)}{2\qty(\rho_1+\rho_2)^2}$\\
$n$ &$\rho_2 - \rho_1$\\
$v$ & $\frac{p_2-p_1}{2}$\\
$b$ & $2\,\qty(\frac{q_2}{\rho_2}-\frac{q_1}{\rho_1})$ \\
$\theta$ & $\frac{\qty(\rho_1+\rho_2) \qty(\phi_1-\phi_2+\pi)+\qty(p_2-p_1)\qty(q_1+q_2)}{\rho_1+\rho_2}$ \\
 \hline
\end{tabular}
\caption{Center-of-mass (CoM) and relative parameters in terms of the canonical variables, i.e., the Faddeev-Takhtaan parameters.}
\label{table:COMCanonical}
\end{table}

Namely, for any given CoM or relative parameter $\chi$, we have
\begin{equation}
\expval{\Delta \chi^2} = \expval{\qty(\,\sum_k \left.\pdv{\chi}{\xi_k}\right|_{\xi_k^0}\delta \widehat{\xi}_k)^2\,},
\label{errorprop}
\end{equation}
where each $\xi_k$ is one of the canonical variables $q_i,p_i,\rho_i,$ or $\phi_i$, and the partial derivatives are evaluated at the values $\xi_k^0$, which are given in Table~\ref{table:2sol_initialcond}. The expectation values are taken with respect to the vacuum defining the chosen noise model. For example, the initial fluctuations in CoM position $B$ with the white noise model are calculated as follows. First we compute
\[
 \Delta \widehat{B}=\sum_k \left.\pdv{B}{\xi_k}\right|_{\xi_k^0}\delta \widehat{\xi}_k =\frac{2}{N}\delta \hat{q}_1+\frac{2}{N}\delta \hat{q}_2;
\]
the other partial derivatives are zero, at least upon evaluating them at the initial values in Table~\ref{table:2sol_initialcond}.

Next we compute 
\[
\expval{\Delta B^2}=\frac{4}{N^2}\left(\rule{0pt}{10pt}\expval{\delta q_1^2}+2\expval{\delta q_1 \delta q_2}+\expval{\delta q_2^2}\right)\,,
\]
where we used the fact that $\expval{\delta q_1 \delta q_2}=\expval{\delta q_2 \delta q_1}$ (see the caption of Table~\ref{table:results_2sol_canonical}). Finally, we plug in the values of $\expval{\delta q_1^2}$, $\expval{\delta q_1 \delta q_2}$, and $\expval{\delta q_2^2}$ from Table~\ref{table:results_2sol_canonical}, obtaining $\expval{\Delta B^2}=\frac{16 \pi^2}{3 N^3}$. This is in Gordon units, so in the SI units, it is $\frac{16 \pi^2}{3 N^3}\unitLG^2$. This agrees with the result in Table~I. of \cite{marchukov2020}; indeed, we exactly reproduce all eight entries in that table.

\subsubsection{3-soliton breather}

Because the canonical formalism reduces the computational complexity, it is possible to push beyond the results already obtained in \cite{marchukov2020}. As a demonstration, we also present initial fluctuations for the 3-soliton breather, also with the white-noise model. The 3-soliton breather is created by a 9-fold quench of the coupling constant, and at $t=0$ the mother soliton converts into three daughter solitons with norm ratios $1:3:5$. Non-zero initial values for the 3-soliton parameters are presented in Table \ref{table:3sol_initialcond}, and the results for the initial fluctuations in all parameters and their covariances are presented in Table \ref{table:init_variances_3sol}, where again the covariances for pairs of variables reduce to a single Lagrange bracket, as in the 2-soliton case. We list the numerical factors only, i.e., we have set $N=1$ (this has the practical advantage of speeding up the computations). But the scalings with $N$ can be deduced using the reasoning presented in the paragraph following \eqnm{covars}: the variances and covariances involving two canonical cordinates ($q_j$'s and $\phi_j$'s, e.g., $\expectationvalue{\delta q_2^2}$ and $\expval{\delta\phi_1\delta\phi_2}$ ) get a prefactor of $1/N$, while those involving two canonical momenta ($p_j$'s and $\rho_j$'s) get a prefactor of $N$.

\begin{table}[h] 
\centering
\begin{tabular}{|c|c|c|c|c|c|c|c|c|c|c|c|}
\hline
 Parameter & $\rho_1$ & $\rho_2$ & $\rho_3$& $\phi_2$ &$\phi_3$ \\ \hline
Initial value & $\frac{1}{9}$ & $\frac{3}{9}$ & $\frac{5}{9}$&  $\pi$ &$2\pi$\\
 \hline
\end{tabular}
\caption{Initial values of parameters for the 3-soliton breather. Parameters not listed have initial value zero.}
\label{table:3sol_initialcond}
\end{table}

\begin{table} 
\centering
\begin{tabular}{|c|c|c|c|c|}
\hline
 \makecell{Expectation\\ value} & \makecell{Initial\\ Fluctuation} & Covariance & \makecell{Initial\\ Fluctuation} \\ \hline
$\expval{\delta q_1^2}$ & $\frac{141 \pi ^2}{140}-\frac{7}{2}$ & $\expval{\delta q_1\delta q_2}$ & $\frac{99 \pi ^2}{140}-\frac{51}{16}$\\
$\expval{\delta q_2^2}$  & $\frac{9}{80} \left(35+8 \pi ^2\right)$ & $\expval{\delta q_1\delta q_3}$ & $\frac{5}{112} \left(12 \pi ^2-77\right)$\\
$\expval{\delta q_3^2}$  & $\frac{5}{112} \left(245+24 \pi ^2\right)$ & $\expval{\delta q_2\delta q_3}$ & $\frac{15}{16}+\frac{9 \pi ^2}{14}$ \\
$\expval{\delta p_1^2}$  & $\frac{43}{693}$ & $\expval{\delta p_1\delta p_2}$ & $-\frac{23}{1155}$ \\
$\expval{\delta p_2^2}$  & $\frac{106}{3465}$ & $\expval{\delta p_1\delta p_3}$ & $\frac{1}{693}$\\
$\expval{\delta p_3^2}$ & $\frac{10}{693}$ & $\expval{\delta p_2\delta p_3}$ & $-\frac{2}{231}$\\
$\expval{\delta \rho_1^2}$ & $\frac{47}{315}$ & $\expval{\delta\rho_1\delta\rho_2}$ & $\frac{11}{105}$ \\
$\expval{\delta \rho_2^2}$ & $\frac{2}{15}$ & $\expval{\delta\rho_1\delta\rho_3}$ & $\frac{5}{63}$ \\

$\expval{\delta\rho_3^2}$ & $\frac{10}{63}$ & $\expval{\delta\rho_2\delta\rho_3}$ & $\frac{2}{21}$ \\
$\expval{\delta\phi_1^2}$ & $\frac{13}{2}+\frac{129 \pi ^2}{308}$ & $\expval{\delta\phi_1\delta\phi_2}$ & $-\frac{3 \left(6545+276 \pi ^2\right)}{6160}$\\
$\expval{\delta\phi_2^2}$ & $\frac{61}{16}+\frac{159 \pi ^2}{770}$ & $\expval{\delta\phi_1\delta\phi_3}$ & $\frac{3 \pi ^2}{308}-\frac{1}{16}$ \\
$\expval{\delta\phi_3^2}$ & $\frac{29}{16}+\frac{15 \pi ^2}{154}$ &$\expval{\delta\phi_2\delta\phi_3}$ & $-\frac{21}{16}-\frac{9 \pi ^2}{154}$ \\
& & $\expectationvalue{Q_iP_j}$ & $\frac{i }{2}\delta_{ij}$\\
& & $\expectationvalue{P_iQ_j}$ & -$\frac{i }{2}\delta_{ij}$\\
 \hline
\end{tabular}
\caption{Initial values and initial quantum fluctuations of the three-soliton breather of unit norm ($N=1$), obtained for the white-noise vacuum state. Correlators not listed are zero. The dependance on $N$ can be restored as follows: the variances and covariances involving two canonical cordinates ($q_j$'s and $\phi_j$'s) get a prefactor of $1/N$, those involving two canonical momenta ($p_j$'s and $\rho_j$'s) get a prefactor of $N$, and mixed ones get no prefactor.}
\label{table:init_variances_3sol}
\end{table}

\subsection{A correlated-noise vacuum}
\label{sec:correlated_noise}

As we said above, the model of white-noise vacuum fluctuations is expected to give answers that are only qualitatively correct. To obtain the actual numbers one expects to see in experiments, one needs to use a more realistic noise model. According to what we said in the introductory part of Sec.~\ref{sec:white_noise}, this will necessarily be a model of `correlated' (`colored') noise.

Based on our discussion in Sec.~\ref{subsec:quantum_state_mother_soliton}, the CoM and relative degrees of freedom are decoupled. While the details of the quantum state of the CoM are largely unknown, the quantum state of the relative degrees of freedom is, to an excellent approximation, given by \eqnm{eq:true_quantum_many-body_ground_state}. And it is only the latter that determines the fluctuations of the relative positions, velocities, phases, and norms of the daughter solitons immediately after the quench.

Since we are well within the weakly interacting regime (see the first paragraph of Sec.~\ref{sec:classical_field}), we can use the Bogoliubov theory to characterize the quantum state of the mother soliton.

\newcommand{\CorrlVacKet}{\ket{0_\text{C}}}
\newcommand{\CorrlVacBra}{\bra{0_\text{C}}}
We begin by laying out the relevant results from Sec.~4 of \cite{Castin2009}. As we mentioned above, these results are obtained using the particle-number-conserving (also called the $U(1)$-symmetry-conserving) Bogoliubov approach \cite{Gardiner1997_1414,Castin1998,Sinatra2007_033616}.

Let us introduce the field operator $\hat{b}_k$, which satisfies the commutation relations 
\begin{equation}
 \begin{aligned}
& \comm{\hat{b}_k}{\hat{b}^\dagger_n}= 2 \pi \delta(k-n) \\
& \comm{\hat{b}_k}{\hat{b}_n}=\comm{\hat{b}^\dagger_k}{\hat{b}^\dagger_n} =0.
\end{aligned}
\label{eq:bk_commutation_rels}
\end{equation}
(See also the Supplemental Material for \cite{marchukov2020}, Sec.~V.)

Next, we define the `correlated-noise vacuum' $\CorrlVacKet$ as the unit-norm state annihilated by this field operator: $\hat{b}_k\CorrlVacKet=0$. In what follows, all expectation values (such as $\bra{0}\delta \hat q(x^\prime)\delta \hat p(y^\prime)\ket{0}$, etc.) are taken with respect to the correlated-noise vacuum. In order to compute them, we need to express the fluctuations $\delta \hat{\psi}$ in terms of the $\hat{b}_k$ and $\hat{b}_k^\dagger$ operators: 
\begin{equation}
  \delta \hat{\psi} = \frac{1}{2\pi}\int_{-\infty}^{\infty} \qty(u_k(x)\hat{b}_k + v^*_k(x)\hat{b}^\dagger_k) \dd k.
  \label{psiflux}
\end{equation}
Here $u_k(x)$ and $v_k(x)$ are the `corrected' expressions for the Bogoliubov modes $U_k(x)$ and $V_k(x)$ of the fundamental soliton, at position $x$ and wavenumber $k$:
\begin{equation}
\begin{aligned}
  \ket{u_k} & = \qty(1 - \dyad{\phi_0}{\phi_0} )\ket{U_k} \\
  \ket{v_k} & = \qty(1 - \dyad{\phi_0}{\phi_0} )\ket{V_k},
\end{aligned}
 \label{corrections}
 \end{equation}
where $\ket{\phi_0}$ is the pure single soliton solution. It is the presence of the projectors $\dyad{\phi_0}{\phi_0}$ which ensures that the $U(1)$ symmetry of the problem is preserved \cite{Castin1998}. The Bogoliubov modes $\ket{U_k}$ and $\ket{V_k}$ are known exactly \cite{Kaup1990}:
\begin{equation}
\begin{aligned}
  & U_{k}(x) = e^{iKX} \hspace{11em} \\
  & \hspace{1em} \times \frac{1+(K^{2}-1)\cosh ^{2}{X}+2iK\sinh {X}\cosh {X}}{%
(K-i)^{2}\cosh ^{2}{X}}, \\
 & V_{k}(x)= e^{iKX}\frac{1}{(K-i)^{2}\cosh ^{2}{X}},
\end{aligned}
\label{Bogoliubov_modes}
\end{equation}
where 
\begin{equation}
  X = \frac{m gN}{2 \hbar^2}x\, , \qquad K = \frac{2 \hbar^2}{mgN}k.
\end{equation}

Note that in \cite{marchukov2020}, the analog of equation \eqref{psiflux} (namely, Eq.~(S-28) in the Supplemental Material for that paper) uses the uncorrected expressions for the Bogoliubov modes; in other words, instead of $u_k(x)$ and $v^*_k(x)$, it uses $U_k(x)$ and $V^*_k(x)$, Further details on the role of these corrections are discussed in Appendix \ref{sec:AppendixCorrTerms}.

Next, we follow the same procedure as for the white noise vacuum, where the fluctuations in canonical coordinates are given in \eqref{eq:fluct_cont} and we use the same definitions of $\delta q(x)$ and $\delta p(x)$ as stated in \eqref{lildeltas}. Using the commutation relations \eqref{eq:bk_commutation_rels}, we find 
\begin{equation}
\begin{aligned}
  \expval{\delta\hat{\psi}(x)\delta\hat{\psi}(y)} & = \frac{1}{2\pi} \int_{-\infty}^{\infty} u_k(x)v^*_k(y) \dd k\,, \\
   \expval{\delta\hat{\psi}(x)\delta\hat{\psi}^\dagger(y)} & = \frac{1}{2\pi} \int_{-\infty}^{\infty} u_k(x)u^*_k(y) \dd k\,, \\
   \expval{\delta\hat{\psi}^\dagger(x)\delta\hat{\psi}(y)} & = \frac{1}{2\pi} \int_{-\infty}^{\infty} v_k(x)v^*_k(y) \dd k\,, \\
   \expval{\delta\hat{\psi}^\dagger(x)\delta\hat{\psi}^\dagger(y)} & = \frac{1}{2\pi} \int_{-\infty}^{\infty} v_k(x)u^*_k(y) \dd k \,,
\end{aligned}
\label{corr_correlators}
\end{equation}
so that the expectation values are now triple integrals with 16 terms each. However, the complexity of this calculation can be reduced by defining
\begin{equation}
\begin{aligned}
  F(x,k) &= \pdv{q(x)}{P} \qty( \vphantom{\frac{}{1}} U_k(x)-V_k(x)-2\braket{\phi_0}{U_k}\phi_0) \\
  & -i \pdv{p(x)}{P} \qty( \vphantom{\frac{}{1}} U_k(x)+V_k(x))\,, \\
   F^*(x,k) &= \pdv{q(x)}{P} \qty( \vphantom{\frac{}{1}} U^*_k(x)-V^*_k(x)-2\braket{\phi_0}{U_k}\phi_0) \\
   &+i \pdv{p(x)}{P} \qty( \vphantom{\frac{}{1}} U^*_k(x)+V^*_k(x))\,,
\end{aligned}
   \label{simplyfyingF}
\end{equation}
and exchanging the order of integration. Note the terms $2\braket{\phi_0}{U_k}\phi_0$; these appear because we use the `corrected' expressions for the Bogoliubov modes. Then, $\expectationvalue{\delta\hat Q^2}$ becomes two one-dimensional integrals which can be evaluated analytically using software. Specifically,
\begin{equation}
   \expval{\delta \hat Q^2}= \frac{1}{4 \pi}\int_{-\infty}^{\infty} \abs{F(k)}^2 \dd k\,,
\end{equation}
where
\begin{equation}
  F(k) = \int_{-\infty}^{\infty} F(x,k) \dd x.
\end{equation} 
The variances for $\expval{\delta\hat P^2}$ can be similarly simplified, defining instead
\begin{multline}
  G(x,k) = \pdv{q(x)}{Q} \qty( \vphantom{\frac{}{1}} U_k(x)-V_k(x)-2\braket{\phi_0}{U_k}\phi_0) \\
  -i \pdv{p(x)}{Q} \qty( \vphantom{\frac{}{1}} U_k(x)+V_k(x))
\label{simplifyingG}
\end{multline}
and its complex conjugate. Note again the appearance of the `correction terms' $2\braket{\phi_0}{U_k}\phi_0$. We obtain
\begin{equation}
   \expval{\delta\hat P^2}= \frac{1}{4 \pi}\int_{-\infty}^{\infty} \abs{G(k)}^2 \dd k\,.
\end{equation}
Finally, for the relevant covariances, we have
\begin{equation}
   \expval{\delta Q\delta P}= \frac{1}{4 \pi}\int_{-\infty}^{\infty} F(k)G^*(k) \dd k \,.
\end{equation}

\subsubsection{Results}

The initial conditions for the canonical parameters of both the 2-soliton breather and the 3-soliton breather are still those listed in Table~\ref{table:2sol_initialcond} and Table~\ref{table:3sol_initialcond}, respectively. Results for the initial fluctuations of the canonical parameters are presented in Table~\ref{table:corr_vac_vars_2sol} for the 2-soliton breather, and in Table~\ref{table:3Sol_corr_vac_vars} for the 3-soliton breather. Using the relationships in Table \ref{table:COMCanonical} and \eqnm{errorprop}, our results for the 2-soliton reproduce the initial fluctuations due to correlated noise for the relative coordinates reported in the second row of Table~II of \cite{marchukov2020}. This time, \cite{marchukov2020} reported only numerical results, keeping at most three significant figures, and all we can say is that our analytic answers agree with those results to the number of significant figures given there.

\begin{table} 
\centering
\begin{tabular}{|c|c|c|c|c|}
\hline
 \makecell{Expectation\\ Value} & \makecell{Initial\\Fluctuation} & Covariance & \makecell{Initial\\Fluctuation} \\ \hline
$\expval{\delta q_1^2}$ & $\frac{3}{4}+\frac{\pi ^2}{10}$ &$\expval{\delta q_1\delta q_2}$ & $-\frac{3}{4}-\frac{\pi ^2}{10}$ \\
$\expval{\delta q_2^2}$ & $\frac{3}{4}+\frac{\pi ^2}{10}$ & $\expval{\delta p_1\delta p_2}$ & $-\frac{9}{280}$\\
$\expval{\delta p_1^2}$ & $\frac{3}{280}$ & $\expval{\delta \phi_1 \delta \phi_2}$ & $-\frac{3}{4}-\frac{3 \pi ^2}{70}$\\
$\expval{\delta p_2^2}$ & $\frac{27}{280}$ &$\expval{\delta\rho_1\delta\rho_2}$ & $-\frac{3}{40}$ \\
$\expval{\delta \rho_1^2}$ &$\frac{3}{40}$ && \\
$\expval{\delta \rho_2^2}$ & $\frac{3}{40}$ &&\\
$\expval{\delta\phi_1^2}$ & $\frac{3}{4}+\frac{\pi ^2}{70}$ &&\\
$\expval{\delta\phi_2^2}$ & $\frac{7}{4}+\frac{9 \pi ^2}{70}$ && \\

 \hline
\end{tabular}%
\caption{Initial quantum fluctuations of the $N=1$, two-soliton breather's canonical parameters, obtained for the correlated vacuum state. Covariances not listed are zero. Dependance on $N$ can be easily restored: it is the same as for the white-noise results in Table~\ref{table:results_2sol_canonical}.}
\label{table:corr_vac_vars_2sol}
\end{table}

\begin{table} 
\centering
\begin{tabular}{|c|c|c|c|c|}
\hline
 \makecell{Expectation\\ value} & \makecell{Initial\\ Fluctuation} & Covariance & \makecell{Initial\\ Fluctuation} \\ \hline
$\expval{\delta q_1^2}$ & $\frac{13}{9}+\frac{38 \pi ^2}{105}$ & $\expval{\delta q_1 \delta q_2}$ & $\frac{1}{12}-\frac{\pi ^2}{35}$ \\
$\expval{\delta q_2^2}$ & $\frac{13}{4}+\frac{6 \pi ^2}{35}$ & $\expval{\delta q_1 \delta q_3}$ & $-\frac{55}{36}-\frac{\pi ^2}{3}$ \\
$\expval{\delta q_3^2}$ & $\frac{5}{252} \left(245+24 \pi ^2\right)$ & $\expval{\delta q_2\delta q_3}$ & $-\frac{10}{3}-\frac{\pi ^2}{7}$ \\
$\expval{\delta p_1^2}$ & $\frac{1784}{31185}$ & $\expval{\delta p_1 \delta p_2}$ & $-\frac{628}{31185}$\\
$\expval{\delta p_2^2}$ & $\frac{776}{31185}$ & $\expval{\delta p_1\delta p_3}$  & $\frac{4}{6237}$ \\
$\expval{\delta p_3^2}$ & $\frac{40}{6237}$ & $\expval{\delta p_2\delta p_3}$ & $-\frac{68}{6237}$ \\
$\expval{\delta \rho_1^2}$ & $\frac{152}{2835}$ & $\expval{\delta\rho_1\delta\rho_2}$& $-\frac{4}{945}$ \\
$\expval{\delta \rho_2^2}$ & $\frac{8}{315}$ & $\expval{\delta\rho_1\delta\rho_3}$ & $-\frac{4}{81}$\\
$\expval{\delta \rho_3^2}$ & $\frac{40}{567}$ & $\expval{\delta\rho_2\delta\rho_3}$ & $-\frac{4}{189}$\\
$\expval{\delta \phi_1^2}$ & $\frac{203}{36}+\frac{446 \pi ^2}{1155}$ & $\expval{\delta \phi_1\delta\phi_2}$   & $-\frac{28}{9}-\frac{157 \pi ^2}{1155}$\\
$\expval{\delta \phi_2^2}$ & $\frac{61}{18}+\frac{194 \pi ^2}{1155}$ & $\expval{\delta\phi_1\delta\phi_3}$ & $\frac{5}{36}+\frac{\pi ^2}{231}$\\
$\expval{\delta \phi_3^2}$ & $\frac{25}{18}+\frac{10 \pi ^2}{231}$ & $\expval{\delta\phi_2\delta\phi_3}$ & $ -\frac{10}{9}-\frac{17 \pi ^2}{231}$\\
\hline
\end{tabular}
\caption{Initial quantum fluctuations of the $N=1$ three-soliton breather's canonical parameters, obtained for the correlated vacuum state. Covariances not listed are zero. Dependance on $N$ can be easily restored: it is the same as in the case of white noise. Thus, the variances and covariances involving two canonical cordinates ($q_j$'s and $\phi_j$'s) get a prefactor of $1/N$, while those involving two canonical momenta ($p_j$'s and $\rho_j$'s) get a prefactor of $N$.}
\label{table:3Sol_corr_vac_vars}
\end{table}

\section{\label{sec:conclusion}Conclusion} 

Small quantum excitations of integrable PDEs with known Hamiltonian structure are computable using the canonical bracket formalism developed in this paper. The canonical approach reduces the computational complexity of the required calculations, which allows results which were previously computed numerically to be calculated analytically. Further, it allows for calculations of new results, for larger systems. We have applied the formalism to the problems of 2- and 3-soliton breathers arising from a quench of the coupling constant in the nonlinear Schr\"odinger equation. In future work, we hope to identify a different integrable PDE to which our formalism can be applied.

\makeatletter
\let\oldp@subsection\p@subsection

\gappto\appendix{\renewcommand{\p@subsection}{}}

\makeatother

\appendix 

\section{}
\label{sec:AppendixUnits}

Let us explain how the various versions of the NLSE relate to one another. One can view this as a question of units. However, since questions of units have the remarkable ability to be (all at the same time) essential, trivial, and confusing, we will start by considering rescalings of variables.

\subsection{Rescalings of variables}

Our starting point is \eqnm{classical_NLSE}, the NLSE written in full SI units, and defined on the whole real line:
\begin{equation}
i \hbar \frac{\partial}{\partial t} \Psi = -\frac{\hbar^2}{2 m}\frac{\partial^2}{\partial x^2} \Psi
+ g \left|\Psi\right|^2\Psi
\label{apndx:classical_NLSE}
\end{equation}
with the norm $N = \int_{-\infty}^{+\infty} |\Psi(x,\,t)|^2 \, dx$. Let us introduce the function $\psi(\bar{x},\,\bar{t})=\eta\Psi(\alpha\bar{x},\,\beta\bar{t})$, where $\alpha$, $\beta$, and $\eta$ are as-yet unspecified positive real numbers. We want the norm of $\psi$, $\int_{-\infty}^{+\infty} |\psi(\bar{x},\bar{t})|^2 \, d\bar{x}$, to also be $N$; this gives $\eta=\sqrt{\alpha}$. 

We divide both sides of \eqnm{apndx:classical_NLSE} by $\hbar$, substitute $\Psi(x,\,t)=(1/\sqrt{\alpha})\psi(x/\alpha,\,t/\beta)$, and multiply both sides through by $\beta\sqrt{\alpha}$. We also make the replacements $x/\alpha=\bar{x}$ and $t/\beta=\bar{t}$. We get
\begin{multline}
 i\frac{\partial}{\partial \bar{t}}\psi(\bar{x},\,\bar{t})=-\frac{\hbar}{2m}\frac{\beta}{\alpha^2}\frac{\partial^2}{\partial \bar{x}^2}\psi(\bar{x},\,\bar{t}) \\
 +g\frac{\beta}{\alpha\hbar}\left|\psi(\bar{x},\,\bar{t})\right|^2\psi(\bar{x},\,\bar{t})\,,
 \label{NLSE_rescaled}
\end{multline}
still defined on the whole real line.

The various forms of the NLSE can all be derived from this equation.

\subsection{A parameter-free form of the NLSE (the Gordon form)}

One obvious way to proceed (and a good way to introduce some relevant isssues that will also be important later) would be to demand that $\frac{\hbar}{m}\frac{\beta}{\alpha^2}=1$ and $|g|\frac{\beta}{\alpha\hbar}=1$. This gives $\alpha=\hbar^2/(m|g|)$ and $\beta=\hbar^3/(mg^2)$. Assuming $g<0$, we get
\begin{equation}
 i\frac{\partial}{\partial \bar{t}}\psi(\bar{x},\,\bar{t})=-\frac{1}{2}\frac{\partial^2}{\partial \bar{x}^2}\psi(\bar{x},\,\bar{t}) 
 -\left|\psi(\bar{x},\,\bar{t})\right|^2\psi(\bar{x},\,\bar{t}).
 \label{Gordon_NLSE}
 \end{equation}
This is the form of the NLSE used by Gordon in \cite{Gordon1983}. Let us refer to these values of $\alpha$ and $\beta$ as $\alpha_\text{G}$ and $\beta_\text{G}$. When no confusion is likely to result, we can rename $\bar{x}$ and $\bar{t}$ to $x$ and $t$.

\subsection{Units and dimensional analysis}
\label{sec:dimensional_analysis}

Before discussing the form used by Faddeev and Takhtajan, let's comment on what we have just done from the point of view of dimensional analysis and units. First, $\alpha_\text{G}$ and $\beta_\text{G}$ have dimensions of length and time, respectively, which follows from the dimensions of $\hbar$ ($ML^2/T$, energy $\times$ time) and $g$ ($ML^3/T^2$, energy $\times$ length). Therefore, $\alpha_\text{G}$ and $\beta_\text{G}$ are clearly the natural units of length and time for the NLSE. The Gordon form of NLSE in \eqnm{Gordon_NLSE} is thus a fully nondimensionalized form of the NLSE, obtained by dividing all lengths and times by their natural units. 

In particular, $\bar{x}$ and $\bar{t}$ are the numerical values of $x$ and $t$ when one uses natural units. This further justifies dropping the bars in \eqnm{Gordon_NLSE}

One might ask where the unit of mass is in all of this: given that we had to introduce the rescaled length $\bar{x}=x/\alpha$ and time $\bar{t}=t/\beta$, why did we not have to also introduce an $\bar{m}=m/\zeta$, where $\zeta$ is a natural unit of mass (which, presumably, is $m$)? The answer is that the dimensions of mass cancel out from \eqnm{apndx:classical_NLSE}. To see that, note that after we divide both sides by $\hbar$, the remaing dimensionful parameters are $\hbar/m$ and $g/\hbar$. Their dimensions are, respectively, $L^2/T$ and $L/T$, neither of which involves $M$ (and, of course, the dimensions of $\Psi$ are $1/\sqrt{L}$, due to the normalization).

Now, a more standard approach to deriving the Gordon form of the NLSE would be to say that we are working in a system of units in which $\hbar=m=|g|=1$. This condition does fix all three fundamental units. Moreover, the resulting units of length and time are just as we found above, i.e., $\hbar^2/(m|g|)$ and $\hbar^3/(mg^2)$, while the unit of mass is obviously $m$. So the two approaches are almost equivalent, except that in our original approach, we recognize that the unit of mass actually need not be fixed. 

To say the same thing in yet another way: suppose that the units of length and time are as above, but the unit of mass is left undetermined as $u_M$. Then, after we change to the new system of units, the numerical values of $\hbar$ and $|g|$ will not be completely determined: both will be equal to $m/u_M$. But---again---since $\hbar$ and $g$ only really enter our equations as $\hbar/m$ and $g/\hbar$, this does not matter, and we obtain a fully nondimensionalized version of the NLSE.

Whichever approach we take, we find that the nondimensionalized form of the NLSE contains no adjustable dimensionless parameters (once the norm $N$ is fixed). This could have been predicted by doing actual dimensional analysis with $\hbar$, $m$, and $g$. Consider the product $\hbar^a m^b |g|^c$. Its dimensions are $M^{a+b+c} L^{2 a+3 c} T^{-a-2 c}$. We are interested forming a dimensionless combination, so we want to solve $a+b+c=0$, $2 a+3 c=0$, and $-a-2 c=0$. However, the coefficient matrix of this system,
\begin{equation}
\setlength\arraycolsep{3pt}
\renewcommand{\arraystretch}{0.75}
 \begin{pmatrix}
 1 & 1& 1\\
 2 & 0 & 3\\
 -1 & 0 & -2
 \end{pmatrix}
\label{coefficient_matrix}
\end{equation}
is invertible (its determinant is 1), and so it is not possible to combine $\hbar$, $m$, and $g$ into a nontrivial dimensionless parameter. Thus, the NLSE has nothing analogous to QED's fine-structure constant \footnote{To simplify the discussion, when considering QED, we will assume it is formulated---as is common practice---in a system of units in which even electromagnetic quantities have dimensions that are powers of mass, length, and time only. Almost always, the Heaviside-Lorentz system is used \cite{Peskin1995,Srednicki2007,Sterman1993} (sometimes called the `rationalized Gaussian' system \cite{Carron2015_150601951,Mandl2010}).%
}. Incidentally, this makes it obvious that the condition for being in the regime of weak interactions, which we found in the first three paragraphs of Sec.~\ref{sec:classical_field}, can only depend on $N$: there are simply no other dimensionless parameters.

 Another consequence of this analysis is that every solution of the (SI) NLSE at one value of $g$ is related by a simple rescaling to a solution of the (SI) NLSE at another value of $g$.
 
\nocite{Peskin1995,Srednicki2007,Sterman1993,Carron2015_150601951,Mandl2010}

This is not to say that the fully nondimensionalized form of the NLSE is always to be preferred. For example, any discussion involving a quench in the coupling constant would be much more convoluted. Also, it would be very difficult to discuss the non-interacting limit, $g\to 0$.

\subsection{The Faddeev-Takhtajan form}
\label{Faddeev-Takhtajan_form}

We are now ready to discuss \eqnm{TakEq}, the form of the NLSE used by Faddeev and Takhtajan:
\begin{equation}
  i \pdv{\psi}{t} = - \pdv[2]{\psi}{x} + 2 \kappa \abs{\psi}^2 \psi \,.
  \label{apndx:TakEq}
\end{equation} 

As in the case of Gordon's form, the Faddeev-Takhtajan form is sometimes described as correspoding to a system of units where one fixes the values of $\hbar$, $m$, and $g$, this time so that $\hbar=1$, $m=1/2$, and $g=2\kappa$. This is consistent with both approaches we will present below, though a caveat similar to that stated for the Gordon form applies: it is not actually necessary to set the unit of mass (though it is not detrimental, either).

Our starting point is again \eqnm{NLSE_rescaled}. We see that the Faddeev-Takhtajan form is recovered if we demand that
\begin{align}
 \frac{\hbar}{m}\frac{\beta}{\alpha^2}&=2 & \text{and} && |g|\frac{\beta}{\alpha\hbar}=2|\kappa|\,. \label{FT_units_codition}
\end{align}
But note that we now have three unknowns ($\alpha$, $\beta$, and $\kappa$), and still only two equations. We are thus free to fix one of these three to any value we want. 

\subsubsection{Option 1: fix $\kappa$.}

In this option, which turns out to be the most convenient for our purposes, we can set $|\kappa|$ to be any positive value, independent of $|g|$. Then the solution of the system is $\alpha=|\kappa|\frac{\hbar^2}{m|g|}=|\kappa| \alpha_\text{G}$ and $\beta=2\kappa^2\frac{\hbar^3}{m g^2}=2\kappa^2\beta_\text{G}$. Since we still want $\alpha$ and $\beta$ to be units of length and time, $\kappa$ must be dimensionles.

In this case, the numerical value of $|\kappa|$ is essentially arbitrary---any change in its value can be compensated for by an appropriate change in the units of length and time. This has the following (confusing) consequence: even though $\kappa$ is a dimensionless parameter proportinal to $g$, it is neverthelss \emph{not} a natural measure of the interaction strength. In other words, it is \emph{not} an analog of QED's fine-structre constant, which \emph{is} a natural measure of the strength of electromagnetic interactions. Indeed, we have already seen above that, on dimensional grounds, such a measure is impossible for NLSE. Note that once the values of the electric charge $e$, the speed of light $c$, and $\hbar$ are fixed in QED \footnote{Again assuming that the Heaviside-Lorentz units are used.}, then the value of the fine-structure constant is fixed also, no matter how one chooses the working units of mass, length, and time. In contrast, even if the SI values of $g$, $\hbar$, and $m$ are fixed in the NLSE, one can still set $|\kappa|$ to any value one likes, provided one adjusts one's working units of length and time.

However, having $\kappa$ as a parameter does make it much easier to discuss quenches and the noninteracting limit $g\to 0$ than in the fully nondimensionalized (Gordon) case. After all, if the units of length and time are kept fixed while $g$ is changed by a factor, then $\kappa$ must be changed by the same factor. Indeed, in their book \cite{takhtajan_book2007}, Faddeev and Takhtajan do have occasion to discuss the limit $\kappa\to0$.

\subsubsection{Option 2: fix the unit of length or the unit of time.} Another interesting way to solve equations~\eqref{FT_units_codition} is to fix either $\alpha$ or $\beta$ to $1$, so that the value of $|\kappa|$ must be solved for and cannot be arbitrarily chosen. Let's discuss the case $\alpha=1$, so that $\beta=2m/\hbar$ and $\kappa=mg/\hbar^2$. In this case, $\kappa$ is not arbitrary. However, it is also not dimensionless: its dimensions are $1/L$. 

Let us interpret this procedure as a change of units. In this new system of units, all quantities have dimensions that are powers of length only. To see how this works, note that for any quantity $A$, there are unique $a$ and $b$ such that $A/[\hbar^a (2m/\hbar)^b]$ has no dimensions of mass or time. Here $a$ is determined by the requirement that $\hbar^a$ have the same mass dimension as $A$, so $A/\hbar^a$ has only length and time dimensions. Then $b$ must be such that $(2m/\hbar)^b$ has the same time dimension as $A/\hbar^a$. Now we can state what this change of units does: for every quantity $A$, it replaces it by $A/[\hbar^a (2m/\hbar)^b]$. For example, $t$ is replaced by $t/(2m/\hbar)$, with dimensions $L^2$; $g$ is replaced by $g/[\hbar\,(2m/\hbar)^{-1}]$, which is indeed $2$ times the $\kappa$ we found above. All this is very similar to how natural units are introduced in QED, also so that all quantities have dimensions that are powers of length alone. The only difference is that instead of $A/[\hbar^a (2m/\hbar)^b]$, in QED one uses $A/(\hbar^a c^b)$ (again, this assumes that the Gaussian or the Heaviside-Lorentz system of units is used); see, e.g., pp.~38 and 222 of \cite{Sterman1993}.

\subsubsection{A relation between the Gordon and the Faddeev-Takhtajan forms}

Let $\psi_\text{G}(x,\,t)$ be a solution, of norm $N$, of the Gordon form of the NLSE, \eqnm{Gordon_NLSE}. Then 
$\psi_\text{F}(x,\,t)=\sqrt{|\kappa|}\,\psi_\text{G}(|\kappa|x,\,2\kappa^2t)$ is a solution, also of norm $N$, of the Faddeev-Takhtajan form of the NLSE, \eqnm{apndx:TakEq}.

\section{}
\label{sec:AppendixCorrTerms}

In this appendix, we will discuss the derivation of the `correction terms' $2\braket{\phi_0}{U_k}\phi_0$ in Eqs.~\eqref{simplyfyingF} and \eqref{simplifyingG}, which appear because we use the corrected expressions for the Bogoliubov modes. In the process, we will explain why our results for the 2-soliton breather match those in \cite{marchukov2020} despite the fact that we included the corrections and \cite{marchukov2020} did not.

We first need to calculate the inner products $\braket{\phi_0}{U_k}$ and $\braket{\phi_0}{V_k}$. 

The Bogoliubov-de Gennes equations are 
\begin{align}
\epsilon_k U_k(x) &= 
\qty[-\frac{\hbar^2}{2m}\pdv[2]{x}+2gN\abs{\phi_0(x)}^2-\mu_0]U_k(x) \hspace{2em} \nonumber \\
& \hspace{10em} +gN\phi_0^2V_k(x) \\
-\epsilon_k V_k(x) &= \qty[-\frac{\hbar^2}{2m}\pdv[2]{x}+2gN\abs{\phi_0(x)}^2-\mu_0]V_k(x) \nonumber \\
& \hspace{10em} +gN\phi_0^{*^2}U_k(x)
\,\,,
\label{BdGEqns}
\end{align}
where $\mu_0$ is the chemical potential.

First, taking the inner product with $\ket{\phi_0}$ to obtain $\braket{\phi_0}{U_k}$, we obtain
\begin{multline}\label{firstinprod}
\epsilon_k\braket{\phi_0}{U_k} = \\
\int_{-\infty}^\infty \dd x\, \phi_0 \qty[-\frac{\hbar^2}{2m}\pdv[2]{x}+2gN\abs{\phi_0(x)}^2-\mu_0]U_k(x)\\
+gN\phi_0^3V_k(x).
\end{multline}
The first term of the integrand can be integrated by parts twice. The boundary terms vanish in both cases since in order to be a soliton solution, $\phi_0$ must go to zero at $\pm \infty$. Thus we have
\begin{multline}
\int_{-\infty}^\infty \dd x\,\qty( -\frac{\hbar^2}{2m}\phi_0 \pdv[2]{x}U_k(x)) \\
= \int_{-\infty}^\infty \dd x\,\qty( -\frac{\hbar^2}{2m}\qty(\pdv[2]{x}\phi_0 )U_k(x)) \,.
\end{multline}

But we know that $\phi_0$ is a solution of the NLSE, so we can substitute 
\begin{equation}\label{subs}
 -\frac{\hbar^2}{2m} \pdv[2]{x}\phi_0= \mu_0\phi_0 - gN\abs{\phi_0(x)}^2\phi_0 
\end{equation}
into (\ref{firstinprod}) for the first term. We find
\begin{multline}\label{almost}
\epsilon_k\braket{\phi_0}{U_k} = \\
\int_{-\infty}^\infty \dd x \left[\mu_0\phi_0- gN\abs{\phi_0(x)}^2\phi_0 +2gN\abs{\phi_0(x)}^2\phi_0 
\right. \\
\left. \vphantom{\abs{\phi_0(x)}^2} -\mu_0\phi_0\right] U_k(x) 
+gN\phi_0^3V_k(x)
\,\,.
\end{multline}
When $\phi_0$ is wholly real, as is the case for the soliton breather model, $\abs{\phi_0}^2 = \phi_0^2$ and \eqref{almost} reduces to 
\begin{align}
\label{exhibita}
\epsilon_k\braket{\phi_0}{U_k} = gN\int_{-\infty}^\infty \phi_0^3\qty[U_k(x) + V_k(x)] \dd x
\,\,.
\end{align}
We now repeat the calculation for the inner product with $V_k(x)$. Again, when $\phi_0$ is real, 
\begin{multline}\label{secinprod}
-\epsilon_k\braket{\phi_0}{V_k} = \\ 
\int_{-\infty}^\infty \dd x \phi_0 \left[-\frac{\hbar^2}{2m}\pdv[2]{x}+2gN\abs{\phi_0(x)}^2-\mu_0\right] V_k(x)\\
+gN\phi_0^3U_k(x)
\,\,.
\end{multline}
Again integrate the first term by parts twice, and make the substitution in (\ref{subs}). The expression reduces to 
\begin{align}\label{exhibitb}
-\epsilon_k\braket{\phi_0}{V_k} =gN\int_{-\infty}^\infty \phi_0^3\qty[U_k(x) + V_k(x)] \dd x.
\end{align}
Thus, for a real $\phi_0$, we have demonstrated that 
\begin{align}\label{eqopp}
\braket{\phi_0}{U_k} =-\braket{\phi_0}{V_k}
\,\,,
\end{align}
and so the corrected modes are
\begin{equation}
\begin{aligned}
\ket{u_k} &=\ket{U_k} - \braket{\phi_0}{U_k}\ket{\phi_0} \\
\ket{v_k} &= \ket{V_k} +\braket{\phi_0}{U_k} \ket{\phi_0} \,.
\end{aligned}
\label{corrected}
\end{equation}

Because the corrections to $U_k(x)$ and $V_k(x)$ are equal in magnitude but opposite in sign, when we insert the corrected modes into the expressions in \eqref{corr_correlators} and factor the resulting integrand, we find the correction term is simply
\begin{equation}
  -2\braket{\phi_0}{U_k} \phi_0 \pdv{q(x)}{X},
  \label{caninical_correction}
\end{equation}
where $X$ is either $P$ for $F(x,k)$ or $Q$ for $G(x,k)$. This means that for any fluctuations where $\pdv{q(x)}{X}=0$, there will be no contribution from the correction term. For the two-soliton with the initial conditions listed in Table \ref{table:2sol_initialcond}, $\pdv{q(x)}{\phi_i} = \pdv{q(x)}{p_i}=0$, so there are no corrections to any $\rho_i$ or $q_i$ fluctuations. However, $\pdv{q(x)}{q_i}$ and $\pdv{q(x)}{\rho_i}$ are not zero. 

Nevertheless, when we actually go through with the computations, for both $\expval{p_i^2}$ and $\expval{p_1 p_2}$ we find that the result does not depend on whether corrected or uncorrected modes are used---despite the fact that $\pdv{q(x)}{q_i}$ is not zero in those cases. The reason for this is as yet unclear. 

Interestingly, the differences in fluctuations in $\phi_1$ and $\phi_2$ calculated using corrected modes compensate for one another: when considering fluctuations in the relative coordinate, there is zero net difference between using the corrected and the uncorrected modes. When using the corrected modes, the fluctuation for $\phi_1$ is increased by $\rho_2-\rho_1$ relative to the result using the uncorrected mode, while the fluctuation for $\phi_2$, relative to the result from the uncorrected mode, decreases by the same amount.

\bibliography{SolitonFlux_04,qfluc_draft_v26,Nonlinear_PDEs_and_SUSY_v040,MyPubs}

\end{document}